\documentclass[journal]{IEEEtran}
\ifCLASSINFOpdf
\else
\fi

\usepackage{bbm}
\usepackage{verbatim}
\usepackage{graphicx}
\usepackage{cite}
\usepackage{url}
\usepackage[cmex10]{amsmath}
\usepackage{amssymb}    
\usepackage{amsmath}
\usepackage{amsthm}
\usepackage{algorithm, algorithmicx, algpseudocode}
\usepackage[caption=false,font=footnotesize]{subfig}
\usepackage{color}
\usepackage{cite}
\usepackage{epstopdf}
\usepackage{calc}
\usepackage{array}
\usepackage{stfloats}
\usepackage{siunitx}
\usepackage{mathtools}
\usepackage{optidef}
\usepackage{textcomp, gensymb}
\usepackage[printonlyused]{acronym}
\usepackage{xcolor}
\usepackage{cleveref}
\usepackage{multirow}
\usepackage{xcolor}
\usepackage{threeparttable}
\usepackage{todonotes}
\usepackage{makecell}

\acrodef{mmW}{millimeter-wave}
\acrodef{5G}{fifth generation}
\acrodef{SSP}{spatial signal processing}
\acrodef{ULA}{uniform linear array}
\acrodef{sub-THz}{sub-teraherz}
\acrodef{LWA}{leaky wave antenna}
\acrodef{BS}{base station}
\acrodef{UE}{user equipment}
\acrodef{SOTA}{state of the art}
\acrodef{AoA}{angle-of-arrival}
\acrodef{AoD}{angle-of-departure}
\acrodef{AWV}{antenna weight vector}
\acrodef{ADC}{analog-to-digital converter}
\acrodef{AGC}{automatic gain control}
\acrodef{AP}{access point}
\acrodef{BB}{baseband}
\acrodef{RSRP}{reference signal received power}
\acrodef{CSI}{channel state information}
\acrodef{COTS}{commercial-off-the-shelf}
\acrodef{PAA}{phased antenna array}
\acrodef{TTD}{true-time-delay}
\acrodef{LoS}{line-of-sight}
\acrodef{NLoS}{non-line-of-sight}
\acrodef{IA}{initial access}
\acrodef{DFT}{discrete Fourier transform}
\acrodef{UDN}{ultra-dense networks}
\acrodef{RF}{radio frequency}
\acrodef{MPC}{multipath component}
\acrodef{BF}{beamforming}
\acrodef{SNR}{signal-to-noise ratio}
\acrodef{SINR}{signal-to-interference-plus-noise ratio}
\acrodef{OFDM}{orthogonal frequency-division multiplexing}
\acrodef{PAPR}{peak-to-average power ratio}
\acrodef{BPSK}{binary phase shift keying}
\acrodef{QPSK}{quadrature phase shift keying}
\acrodef{ML}{maximum likelihood}
\acrodef{DSP}{digital signal processing}
\acrodef{LUT}{lookup table}
\acrodef{MIMO}{multiple-input multiple-output}
\acrodef{IC}{integrated circuits}
\acrodef{PS}{phase shifter}
\acrodef{DAC}{digital-to-analog converter}
\acrodef{EVM}{error vector magnitude}
\acrodef{CP}{cyclic prefix}
\acrodef{FPGA}{field programmable gate arrays}
\acrodef{MSE}{mean squared error}
\acrodef{RMSE}{root mean square error}
\acrodef{MMSE}{minimum mean square error}
\acrodef{VTC}{voltage-to-time converter}
\acrodef{TDC}{time-to-digital converter}
\acrodef{CMOS}{complementary metal–oxide–semiconductor}
\acrodef{MUX}{multiplexer}
\acrodef{MAC}{multiply-and-accumulate}
\acrodef{CLK}{clock}
\acrodef{PI}{phase interpolator}
\acrodef{FoM}{figure-of-merit}
\acrodef{HI}{hardware impairment}
\acrodef{CS}{compressive sensing}
\acrodef{RST}{reset}
\acrodef{PM}{phase margin}
\acrodef{SCA}{switched-capacitor arrays}
\acrodef{OTA}{operational transconductance amplifier}
\acrodef{LO}{local oscillator}
\acrodef{SS}{Synchronization Signal}
\acrodef{PSS}{Primary Synchronization Signal}
\acrodef{SSS}{Secondary Synchronization Signal}
\acrodef{SSP}{Spatial Signal Processing}
\acrodef{DMRS}{DeModulation Reference Signal}

\DeclareMathOperator*{\argmax}{argmax}

\newcommand{\BW}[0]{\mathrm{BW}}

\newcommand{\sigmaN}[0]{\sigma_{\text{N}}}

\newcommand{\fc}[0]{f_{\text{c}}}

\newcommand{\FBW}[0]{\text{FBW}_{\text{3dB}}}

\newcommand{\T}[0]{\text{T}}
\newcommand{\R}[0]{\text{R}}

\newcommand{\hermitian}[0]{\text{H}}

\newcommand{\tot}[0]{\text{tot}}

\newcommand{\rot}[0]{\text{rot}}

\begin{document}

%
\title{Wideband Beamforming with Rainbow Beam Training using Reconfigurable True-Time-Delay Arrays for Millimeter-Wave Wireless }

%

\author{Chung-Ching~Lin,~\IEEEmembership{Graduate Student~Member,~IEEE,} 
        Veljko~Boljanovic,~\IEEEmembership{Graduate Student~Member,~IEEE,}
        Han~Yan,~\IEEEmembership{Graduate~Student~Member,~IEEE,}        
        Erfan Ghaderi,~\IEEEmembership{Member,~IEEE,} 
        Mohammad~Ali~Mokri,~\IEEEmembership{Graduate Student~Member,~IEEE,}
        Jayce~Jeron~Gaddis,~\IEEEmembership{Student~Member,~IEEE}
        Aditya~Wadaskar,~\IEEEmembership{Graduate~Student~Member,~IEEE,}
        Chase Puglisi,~\IEEEmembership{Member,~IEEE,} 
        Soumen~Mohapatra,~\IEEEmembership{Graduate~Student~Member,~IEEE,}
        Qiuyan~Xu,~\IEEEmembership{Graduate~Student~Member,~IEEE,}
        Sreeni~Poolakkal,~\IEEEmembership{Graduate~Student~Member,~IEEE,}
        Deukhyoun~Heo,~\IEEEmembership{Senior~Member,~IEEE,}
        Subhanshu~Gupta,~\IEEEmembership{Senior~Member,~IEEE,}
        and~Danijela~Cabric,~\IEEEmembership{Fellow,~IEEE}
        
\thanks{This work was supported in part by NSF under grants 1955672, 1705026, and 1944688. This work was also supported in part by the ComSenTer and CONIX Research Centers, two of six centers in JUMP, a Semiconductor Research Corporation (SRC) program sponsored by DARPA.}%
\thanks{Veljko Boljanovic, Han Yan, Aditya Wadaskar, and Danijela Cabric are with the Department of Electrical and Computer Engineering, University of California, Los Angeles, Los Angeles, CA 90095 USA (e-mail:
vboljanovic@ucla.edu).}
\thanks{Chung-Ching Lin, Erfan Ghaderi, Chase Puglisi, Soumen Mohapatra, Mohammad Ali Mokri, Deukhyoun Heo, and Subhanshu Gupta are with the School of Electrical Engineering and Computer Science, Washington State University, Pullman, WA 99164 USA (e-mail: chung-ching.lin@wsu.edu).}
}

\markboth{}%
{}
%



\maketitle

\begin{abstract} 
The decadal research in integrated true-time-delay arrays have seen organic growth enabling realization of wideband beamformers for large arrays with wide aperture widths. This article introduces highly reconfigurable delay elements implementable at analog or digital baseband that enables multiple \ac{SSP} functions including wideband beamforming, wideband interference cancellation, and fast beam training. Details of the beam-training algorithm, system design considerations, system architecture and circuits with large delay range-to-resolution ratios are presented leveraging integrated delay compensation techniques. The article lays out the framework for true-time-delay based arrays in next-generation network infrastructure supporting 3D beam training in planar arrays, low latency massive multiple access, and emerging wireless communications standards. \end{abstract}


%
\IEEEpeerreviewmaketitle

%
%

\section{Introduction}
\label{sec:introduction}
%
%
%
%
\IEEEPARstart{W}{ireless} networks have fueled socio-economic growth worldwide and are expected to further advance to enable new applications such as autonomous vehicles, virtual and augmented reality, and smart cities. Due to shortage of sub-\SI{6}{\giga\hertz} spectrum, \ac{mmW} frequencies play an important role in the \ac{5G} communication networks. Recent research and development of \ac{5G} \ac{mmW} networks has revealed that the propagation loss in the \ac{mmW} band \cite{6824746} needs to be compensated by antenna array gain \cite{Rappaport:mmWavewillwork} and densification of base stations with cell radius as small as a hundred meters \cite{7010535,6736747}. To make radio chipsets power and cost- efficient, \ac{SOTA} \ac{5G} \ac{mmW} transceivers are designed with \ac{PAA} based subarray architecture \cite{8255763,8316768}. As a consequence, signal processing techniques \cite{mmW_SP} and network protocols \cite{mmW_MAC} for \ac{5G} \ac{mmW} networks are designed under constraints of \ac{PAA} architectures. 
\begin{figure}[t]
    \centering
	\vspace{0mm}
	\includegraphics[width=0.5\textwidth]{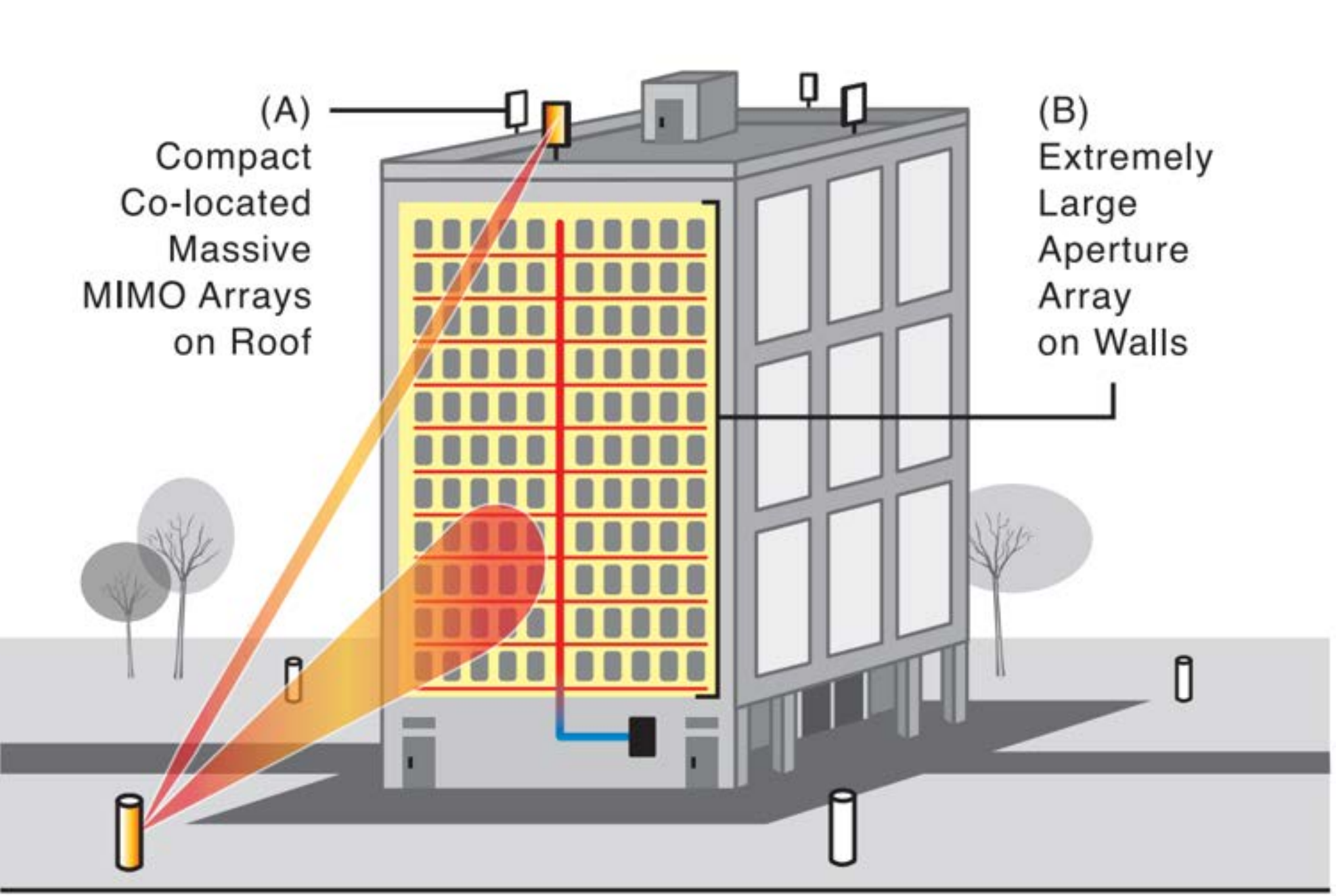}
	\vspace{-4mm}
	\caption{ \color{black} Emerging large multi-antenna based beamforming infrastructure array systems feeding to a baseband \ac{SSP} unit.}
	\vspace{-4mm}
	\label{fig:largearray}
\end{figure}

Future generations of \ac{mmW} networks will operate in the upper \ac{mmW} frequency band where more than \SI{10}{\giga\hertz} bandwidth can be used to meet the ever-increasing demands \cite{8782879,8766143}. Their realization will demand addressing a completely new set of challenges including wider bandwidths, larger antenna array size, and higher cell density at the physical infrastructure level as highlighted in Fig.~\ref{fig:largearray}. These new system requirements demand fundamental rethinking of radio architectures, signal processing and networking protocols. Major breakthroughs are thus required in radio front-end architectures to enable wideband \ac{mmW} networks, as most commonly adopted \ac{PAA} based radios face many challenges in meeting the demanding requirements for different \ac{SSP} functions.

A large portion of \ac{PAA} implementations approximate the inter-element delay between the received signals with a phase-shift and hence the spatial processing is performed based on the phase difference between the received signals. This approximation simplifies the physical integrated circuit implementation, as compensating a phase shift is much simpler than a time delay. However, this simplification and approximation comes at the cost of limited operating fractional bandwidth of the receiver and will be analyzed further in Section~\ref{sec:beamsquint}. 


%

This article brings forward advances in \ac{TTD} arrays that have the potential to significantly impact \ac{SSP} for emerging wireless communication standards. We use \ac{TTD}-based \ac{PAA} \cite{rotman2016a,chu2007,hashemi2008a,chu2013a,garakoui2015a,hu2015a,cho2018,jang2018,mondal2018,Jang2019,dastjerdi2019,ghaderi2019a,ghaderi2020,spoof2020,Ghaderi2021,Lin2021} to first establish a contrast with the current phase shift only \ac{PAA}s  \cite{paramesh2004,hajimiri2005,natarajan2005,natarajan2006,jeon2008,krishnaswamy2010,valdes2010,soer2011,natarajan2011,soer2012,ghaffari2014,krishnaswamy2016,zhang2016,zhang2017,huang2017,johnson2019,huang2019,huang2019jssc,mondal2019} that will be used for both data communications and direction finding. The application of \ac{TTD}-based \ac{PAA}s for different \ac{SSP} functions will have ramifications for future \ac{mmW} network infrastructure and emerging wireless standards. We next develop a \ac{TTD}-based \ac{PAA} exploiting the so called \textit{beam squint} phenomenon to achieve precision direction finding overcoming fundamental latency bottlenecks in earlier beam training methods.

%
%

\section{Exploiting Delay Compensation in Spatial Signal Processors} 
\label{sec:beamsquint}

\begin{figure}[t]
    \begin{center}
        \includegraphics[width=0.3\textwidth]{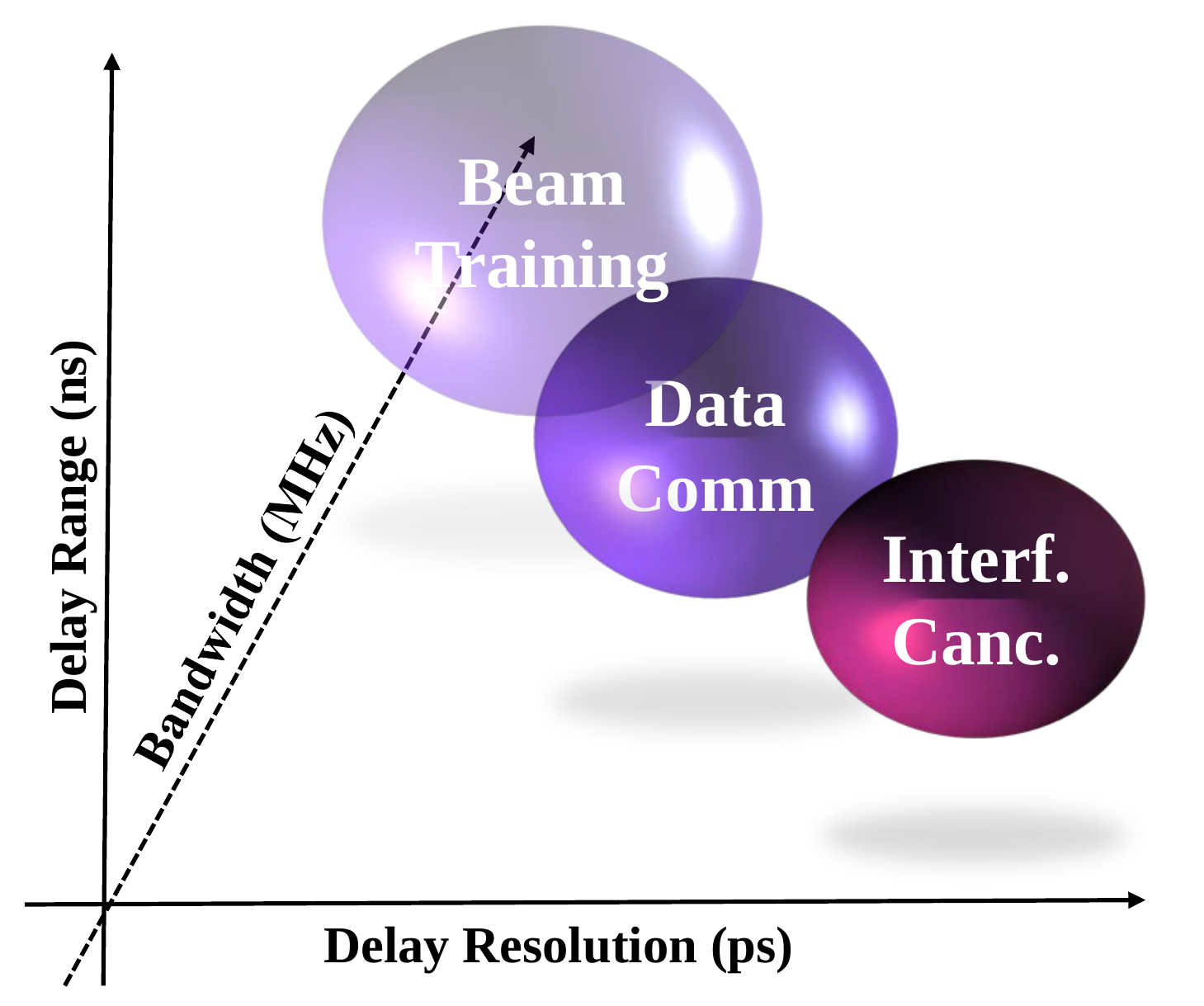}
    \end{center}
    \vspace{-4mm}
    \caption{Trade-offs in different \ac{SSP} modes: data communications, beam-training, and independent interference cancellation. }
    \vspace{-4mm}
    \label{fig:sspmodes}
\end{figure}

\ac{PAA} operating over a very wide frequency band exhibits frequency-dependent beam pattern in an uncontrollable manner and it often degrades beamforming gain and directionality of the beam. This phenomenon is referred as \textit{spatial wideband effect} \cite{8354789} or \textit{beam squint} \cite{cai2018modeling,Gao_squint_magazine,8454468}, that becomes more significant in large antenna arrays. 

This section will describe recent advances in \ac{SSP} algorithms and architectures leveraging delay compensating circuits to overcome fundamental limits in analog \ac{PAA}s. Interested readers are referred to \cite{Boljanovic2021} for a similar analysis for hybrid \ac{TTD} \ac{PAA}s. We will describe application of \ac{TTD} arrays to realize different \ac{SSP} modes when the array is used for data communications or for direction finding. In the former \ac{SSP} mode, the \ac{TTD} array will alleviate beam squint effects while in the latter mode, intentional beam squint is introduced to achieve accurate but fast beam-training significantly reducing the search latency in existing analog \ac{PAA}s. To realize these different \ac{SSP} modes in a \ac{TTD} \ac{PAA}, it is important to understand the trade-offs between the delay range, delay resolution, and modulated bandwidth as shown in Fig.~\ref{fig:sspmodes}. Higher delay resolutions and moderate bandwidth are needed for beam-nulling (independent interference cancellation \cite{ghaderi2019b}) whereas large delay range, moderate resolution, and large bandwidths are needed for beam-training. The \ac{SSP} mode for independent interference cancellation is a corollary of the beamforming \ac{SSP} and thus will not be studied in this article. 

\begin{figure}[t]
    \centering
	\vspace{0mm}
	\includegraphics[width=0.25\textwidth]{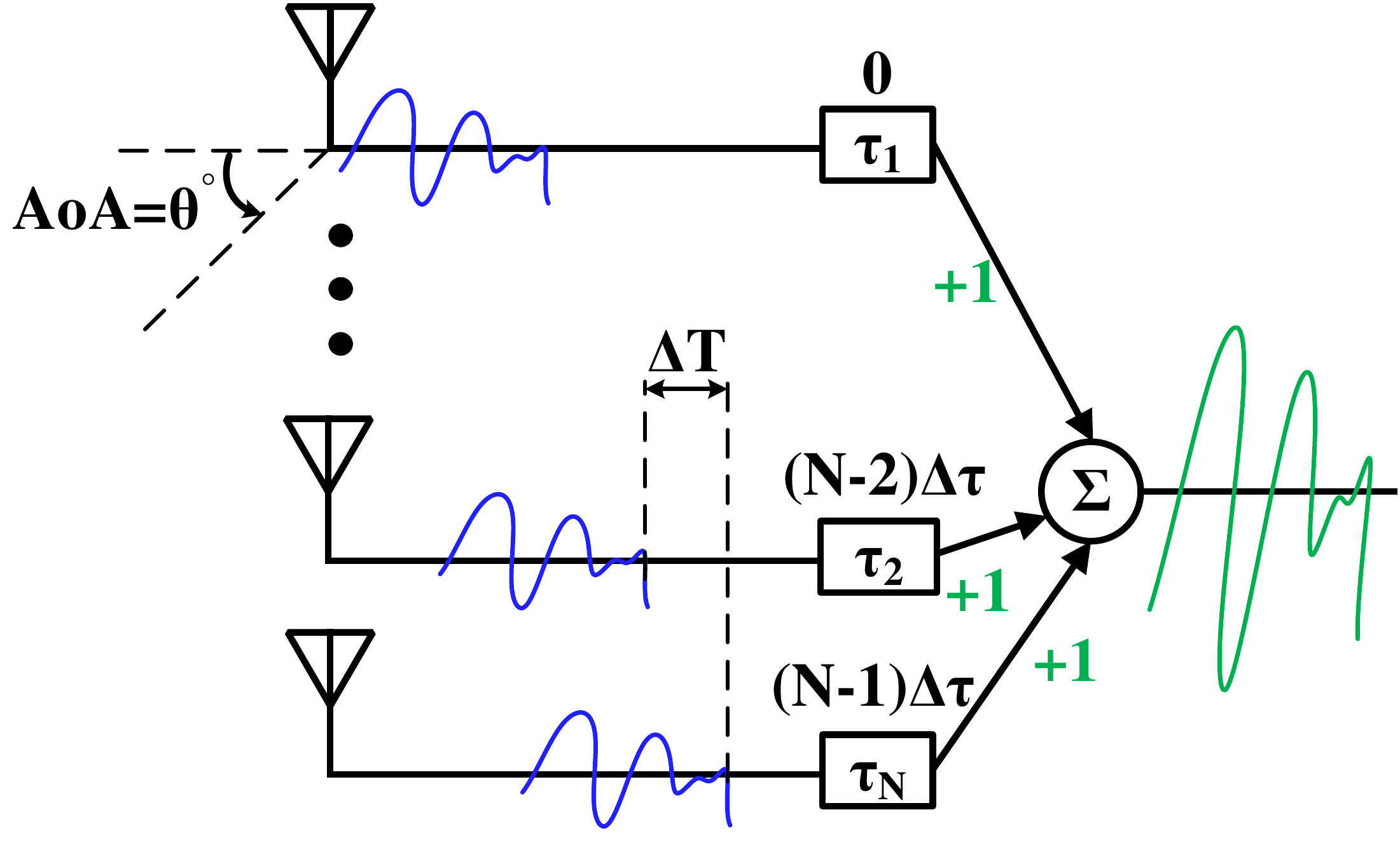}
	\vspace{-4mm}
	\caption{ \color{black} Beamforming in an $N$-element linear array with \ac{TTD} elements.}
	\vspace{-2mm}
	\label{fig:ttd_array_ideal}
\end{figure}
%

\subsection{Mitigating beam-squint for data communications mode}

\begin{figure}[t]
    \centering
	\vspace{-2mm}
	\includegraphics[width=0.25\textwidth]{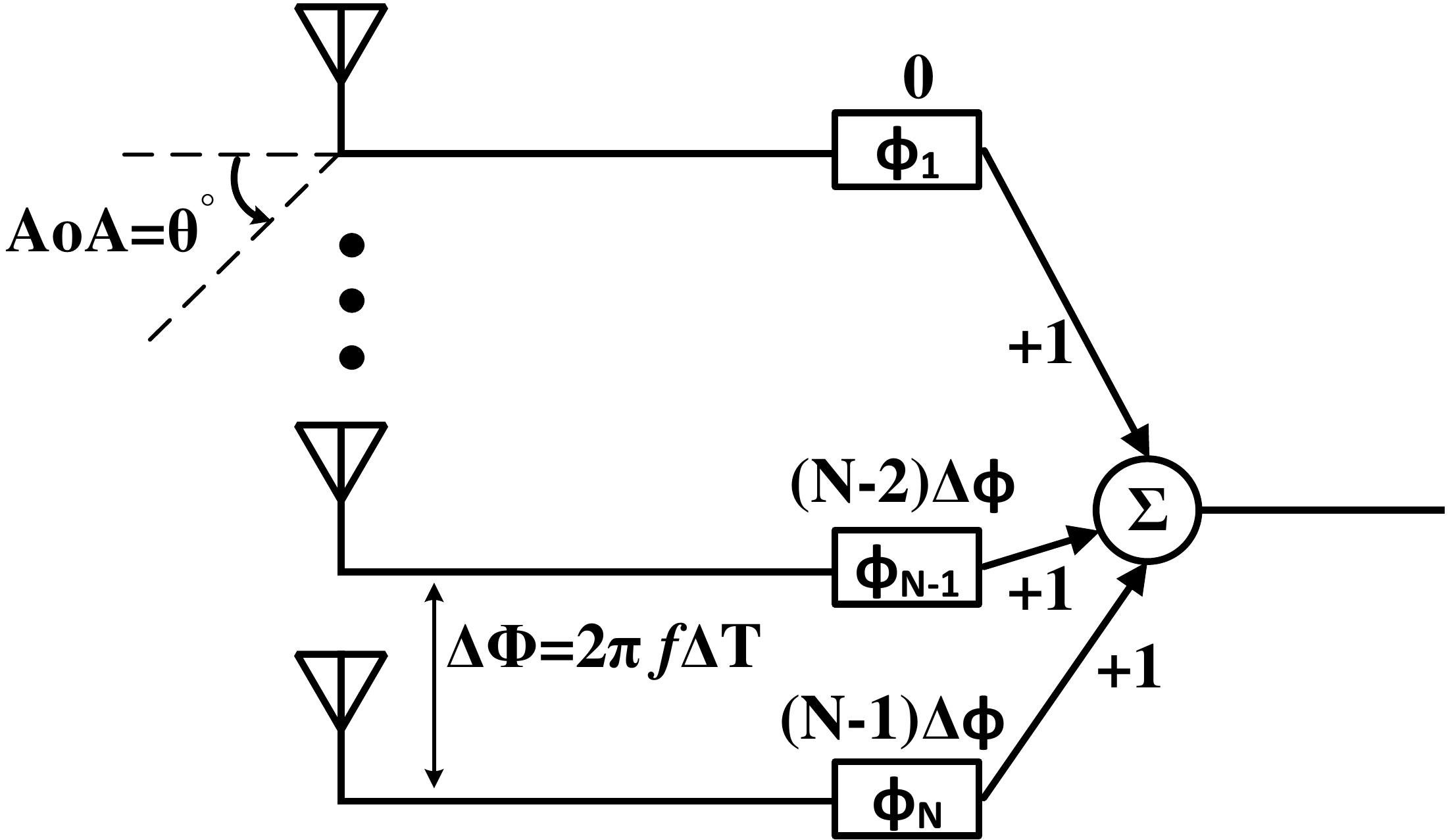}
	\vspace{-2mm}
	\caption{ \color{black} Generalized \ac{PS}-based implementation in an $N$-element \ac{PAA}.}
	\vspace{-3mm}
	\label{fig:ps_array_ideal}
\end{figure}

To alleviate the beam squint effect, the \ac{PAA} is required to compensate inter-element signal delays before any \ac{SSP} is done as illustrated in Fig.~\ref{fig:ttd_array_ideal}. Assuming a half-wavelength antenna spacing, the delay difference between the received signals in two consecutive antennas $\Delta T$ can thus be modeled as $\Delta T=d \sin(\theta)/c$, which can be also expressed as a frequency-dependent phase shift difference $\Delta\Phi$ in the frequency domain through the following expression:
\begin{equation}
         \Delta\Phi=2\pi  f  \Delta T = \pi \sin(\theta)\frac{f}{\fc}
         \label{eq:signal_phase_difference}
\end{equation}
where $f$ is frequency. If the signal of interest is narrowband, i.e., $f/\fc\approx 1$, the required \ac{TTD} element can be replaced with a frequency-flat \ac{PS}. This approximation is the basis of large portion of the \ac{SOTA} \ac{SSP} systems \cite{hajimiri2005,natarajan2005,natarajan2006,jeon2008,krishnaswamy2010,valdes2010,natarajan2011,johnson2019,krishnaswamy2016,zhang2016,zhang2017,huang2017,huang2019,huang2019jssc,soer2011,soer2012,ghaffari2014,paramesh2004,mondal2018,mondal2019}. A generalized implementation of phase shifting based \ac{SSP} unit is shown in Fig.~\ref{fig:ps_array_ideal}.

\begin{figure}[t]
	\vspace{0mm}
	\centering
	\includegraphics[width=0.4\textwidth]{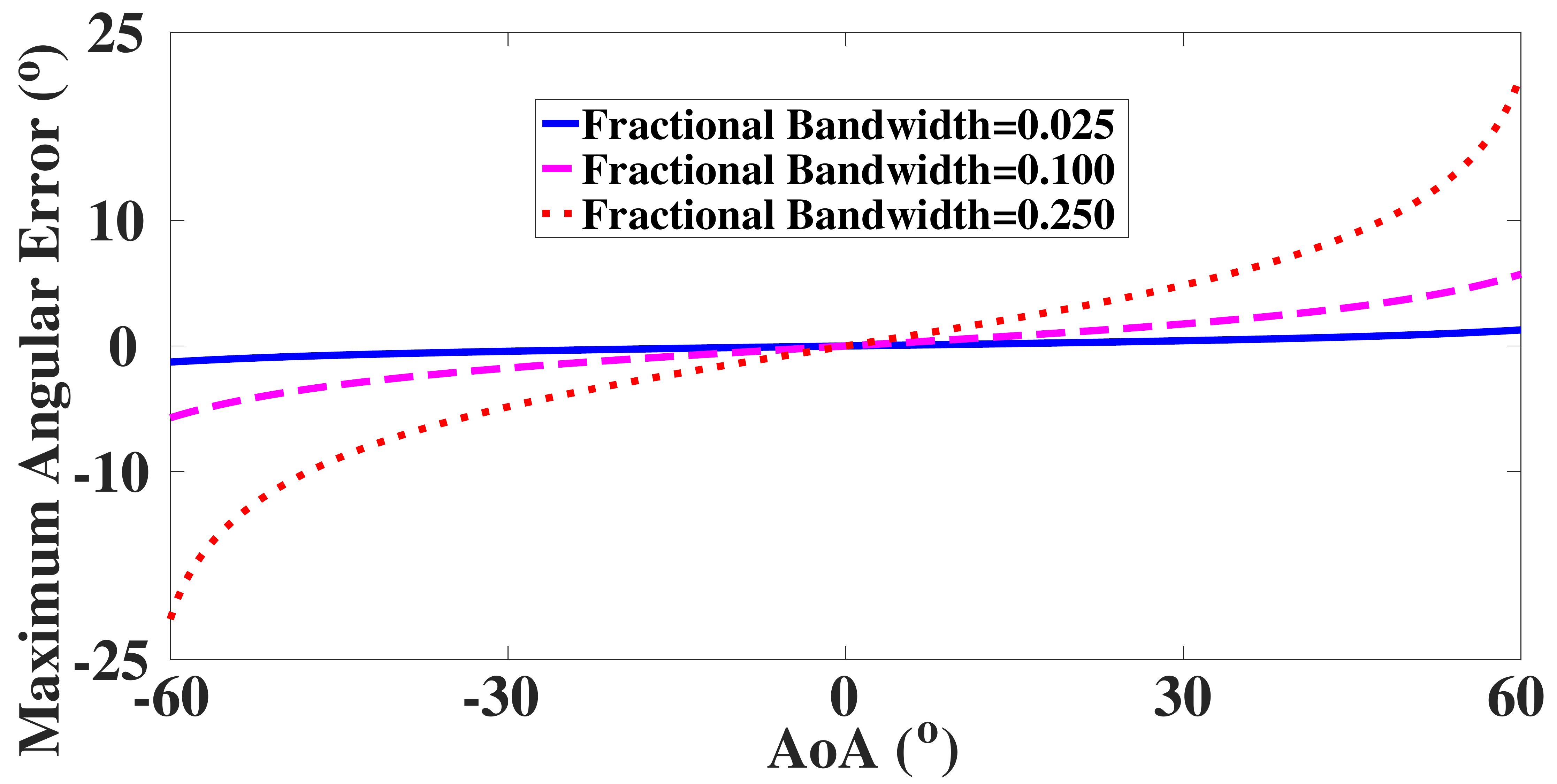}  
	\vspace{-2mm}
	\caption{ \color{black} \ac{PS}-based implementation maximum angular error versus actual \ac{AoA}.}
	\vspace{-4mm}
	\label{fig:PS_FBW}
\end{figure}

As for any approximation, replacing a \ac{TTD} element with a \ac{PS} imposes performance limitations. These limitations can be described both in the angular domain and the frequency domain. In the angular domain, \ac{PS}-based implementation causes the intended angle-of-arrival (\ac{AoA}) to be frequency-dependent. To be more specific, the intended \ac{AoA}, for the both beamforming and beam-nulling cases, varies with the signal of interest frequency as in:

\begin{equation}
    \theta=\sin^{-1}\left(\frac{\fc}{f}\frac{\Delta\Phi}{\pi}\right)
\end{equation}
At the center frequency of the band, the \ac{PS}-based and the \ac{TTD}-based \ac{AoA}s are the same and the beam is formed/nulled. As the frequency moves away from the center, the approximation of \ac{TTD} with a \ac{PS} becomes less valid and there will be an error between the actual and real intended \ac{AoA}. This error can be formulated as follows:
\begin{equation}
    \Delta\theta=\left|\sin^{-1}\left(\frac{\fc}{f}\frac{\Delta\Phi}{\pi}\right) - \sin^{-1}\left(\frac{\Delta\Phi}{\pi}\right)\right|
\end{equation}
This error depends on the actual \ac{AoA} and the relative frequency of the interest. The maximum error happens at the edges of the frequency band, as the frequency deviates the most from the center frequency. As the error is inversely proportional to the frequency, at the lower side of the frequency band error will be larger than the higher side of the band and this maximum error can be found as: 
\begin{align}
    \Delta\theta_{\max} &= \left|\sin^{-1}\left(\frac{\fc}{\fc-\BW/2}\frac{\Delta\Phi}{\pi}\right) - \sin^{-1}\left(\frac{\Delta\Phi}{\pi}\right)\right| 
\end{align}
where $\BW$ is the signal bandwidth and $\BW/\fc$ is called the fractional bandwidth. In Fig.~\ref{fig:PS_FBW}, the maximum error in the angular domain for a \ac{PS}-based \ac{SSP} unit is plotted versus the ideal intended \ac{AoA}, for three cases of fractional bandwidths. In this plot the \ac{AoA} range is limited to $\pm60^{\circ}$, since for larger \ac{AoA}s the actual \ac{AoA} can be as high as $\pm90^{\circ}$ and the error in those cases does not reflect the severity of the \ac{PS}-based implementation. As it can be seen, this error can get as high as $22^{\circ}$ which results in non-alignment with the intended transmitter and consequently loss in the desired signal (beamforming case) or imperfect cancellation (beam-nulling case). The frequency-dependent approximation of a \ac{TTD} element with a \ac{PS} results in frequency-dependent beamforming gain that acts as a bandpass filter \cite{Jang2019,ghaderi2019a}, and imperfect frequency-dependent beam-nulling that results in wideband interference leakage \cite{ghaderi2019b,ghaderi2020}. 

The beamforming gain in a \ac{PS}-based $N$-element receiver can be modeled as the absolute value of the inner product $G(f) = \left| \mathbf{w}^{\hermitian} \mathbf{a} \right|$, where $\mathbf{w}$ is a receive beamformer and $\mathbf{a}$ is a spatial response vector. Assuming a phase difference of $\Delta \phi$ between neighboring \ac{PS}s, the $n$-th element of $\mathbf{w}$ is defined as $[\mathbf{w}]_n =\exp (-j(n-1)\Delta\phi) $. On the other hand, using (\ref{eq:signal_phase_difference}), the $n$-th element of $\mathbf{a}$ is defined as $[\mathbf{a}]_n =\exp(-j(n-1)\Delta \Phi) $. Expressed as a sum of complex exponentials, the gain $G(f)$ becomes
\begin{equation}
    G(f)  = \left| \sum_{n=1}^{N} e^{j(n-1)(\Delta\phi-\Delta \Phi)} \right|.
    \label{equation:psgain}
\end{equation}

\begin{figure}[t]
	\vspace{0mm}
	\centering
	\includegraphics[width=0.4\textwidth]{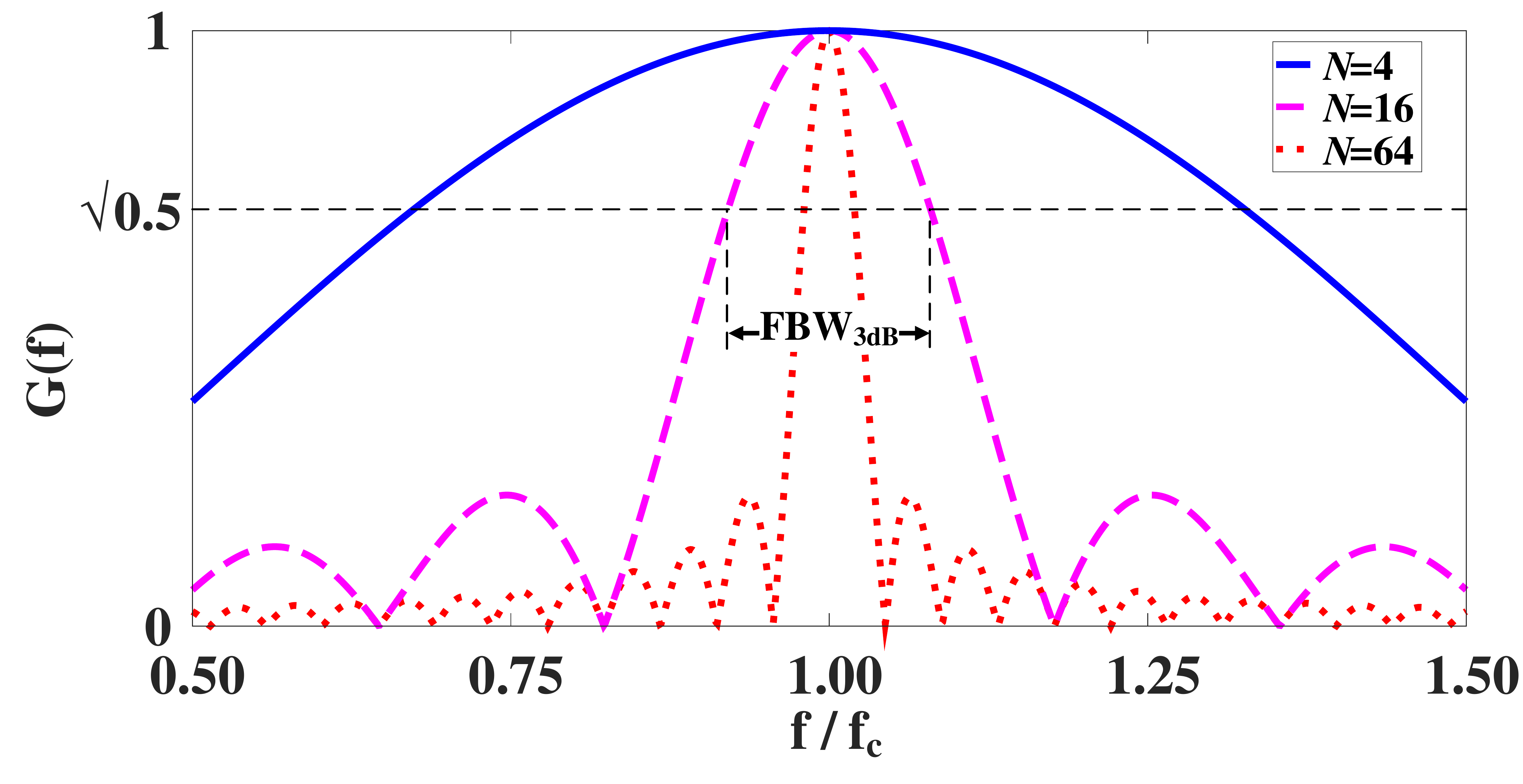}
	\vspace{-4mm}
	\caption{ \color{black} \ac{PS}-based normalized beamforming gain versus normalized frequency.}
	\vspace{-2mm}
	\label{fig:PS_normBW}
\end{figure}

In Fig.~\ref{fig:PS_normBW}, the normalized beamforming gain $G(f)/N$ for intended \ac{AoA} of $45^{\circ}$ versus normalized frequency $f/\fc$ of the input signal for three cases of $N = 4, 16, 64$, are plotted. The beamforming gain in a \ac{PS}-based implementation acts as bandpass filter for the desired signal. Similar to a filter, the 3-dB fractional bandwidth ($\FBW$) can be defined as the fractional bandwidth where the beamforming gain drops 3 dB compared to the maximum value ($N$). As proven in \cite{mailloux_phased_2005,orfanidis2016}, for large values of N, the $\FBW$ is given as:
\begin{equation}
    \FBW \cong \frac{1.772}{N|\sin(\theta)|}
\end{equation}
For larger arrays $\FBW$ becomes smaller and the approximation of a \ac{TTD} element with a \ac{PS} becomes less valid. Similar to the outcome of the beam squinting error, as the intended \ac{AoA} increases, the $\FBW$ gets smaller and the effective error increases.

\begin{figure}[t]
    \centering
    \includegraphics[width=0.45\textwidth]{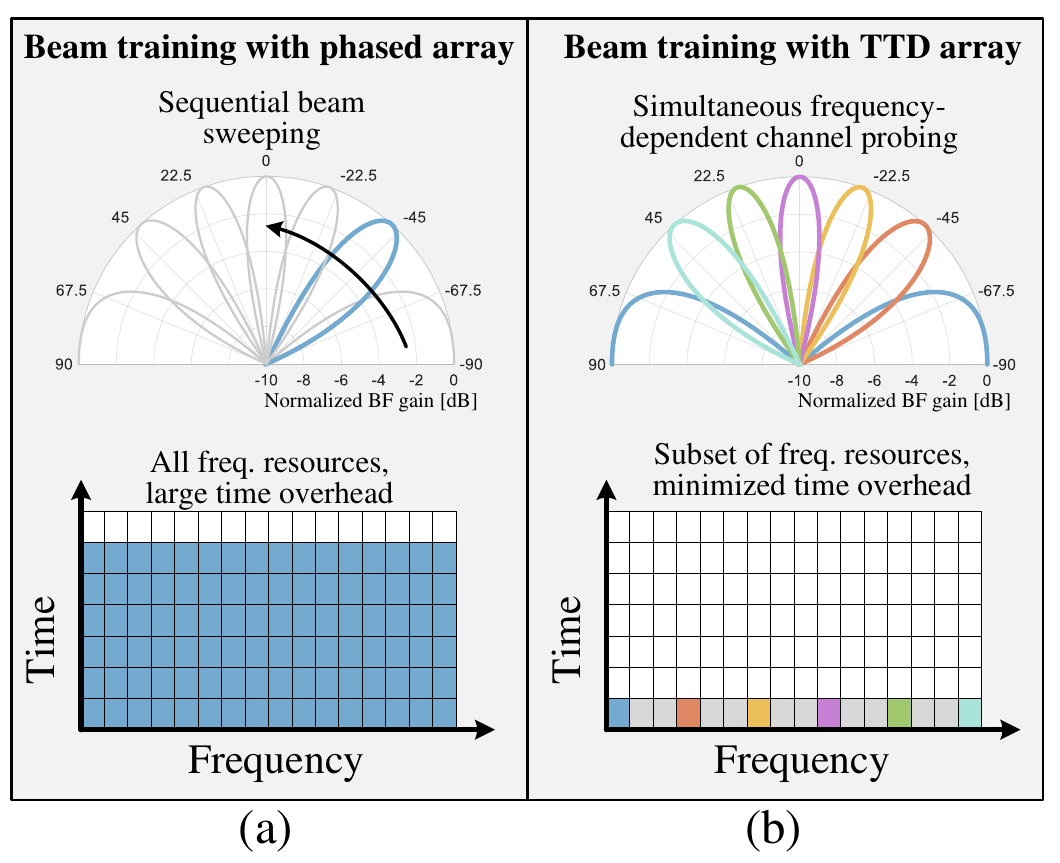}
    \vspace{-2mm}
    \caption{Beam training comparison of (a) \ac{PS} based array with (b) \ac{TTD} array. }
    \vspace{-4mm}
    \label{fig:beam_training}
\end{figure}

\subsection{Beam-squint array for beam-training mode}
Beam-training is a part of \ac{IA} protocol in \ac{mmW} networks that achieve alignment of beamforming directions to realize a maximum gain between \ac{BS} and \ac{UE} \cite{HY_JSTSP,Zorzi:tutorial}. Beam-training using \ac{PAA} requires extensive beam sweeping to estimate \ac{AoD} and \ac{AoA} (\Cref{fig:beam_training}(a)), as it can only probe one steering direction at a time. As a result, \ac{PAA}-based \ac{IA} latency increases with array size and number of users \cite{HY_JSTSP}. The majority of existing \ac{mmW} beam-training algorithms was designed for analog \ac{PAA} due to their power efficiency. Phased arrays use frequency-flat \ac{PS} in all antenna branches to steer/combine the signal in a desired direction. With a single \ac{mmW} chain, \ac{PAA}s can synthesize only one spatial beam with all frequency components being aligned in the same direction. The existing beam-training schemes with \ac{PAA}s include various types of exhaustive beam sweeping where different beam candidates are sequentially used for channel probing to find the \ac{AoA} \cite{Jeong:sweeping2015, Kim:fast, Hosoya:wifi2014, Zhou:efficient}. The required number of probing beams linearly scales with the number of antenna elements $N$, thus, the beam-training overhead becomes a bottleneck for low-latency communications. \ac{SOTA} \ac{IA} approaches based on \ac{5G} requirements are limited to narrowband pilots and relatively small number of antennas, so the \ac{IA} latency is acceptable. However, future evolutions of \ac{mmW}-nets will be \ac{IA} latency limited and would require using wideband pilots and new radio architectures that can probe a large number of \ac{AoA}/\ac{AoD} in parallel as illustrated in \Cref{fig:beam_training}(b).

%

\begin{figure}[t]
	\includegraphics[width=1\columnwidth]{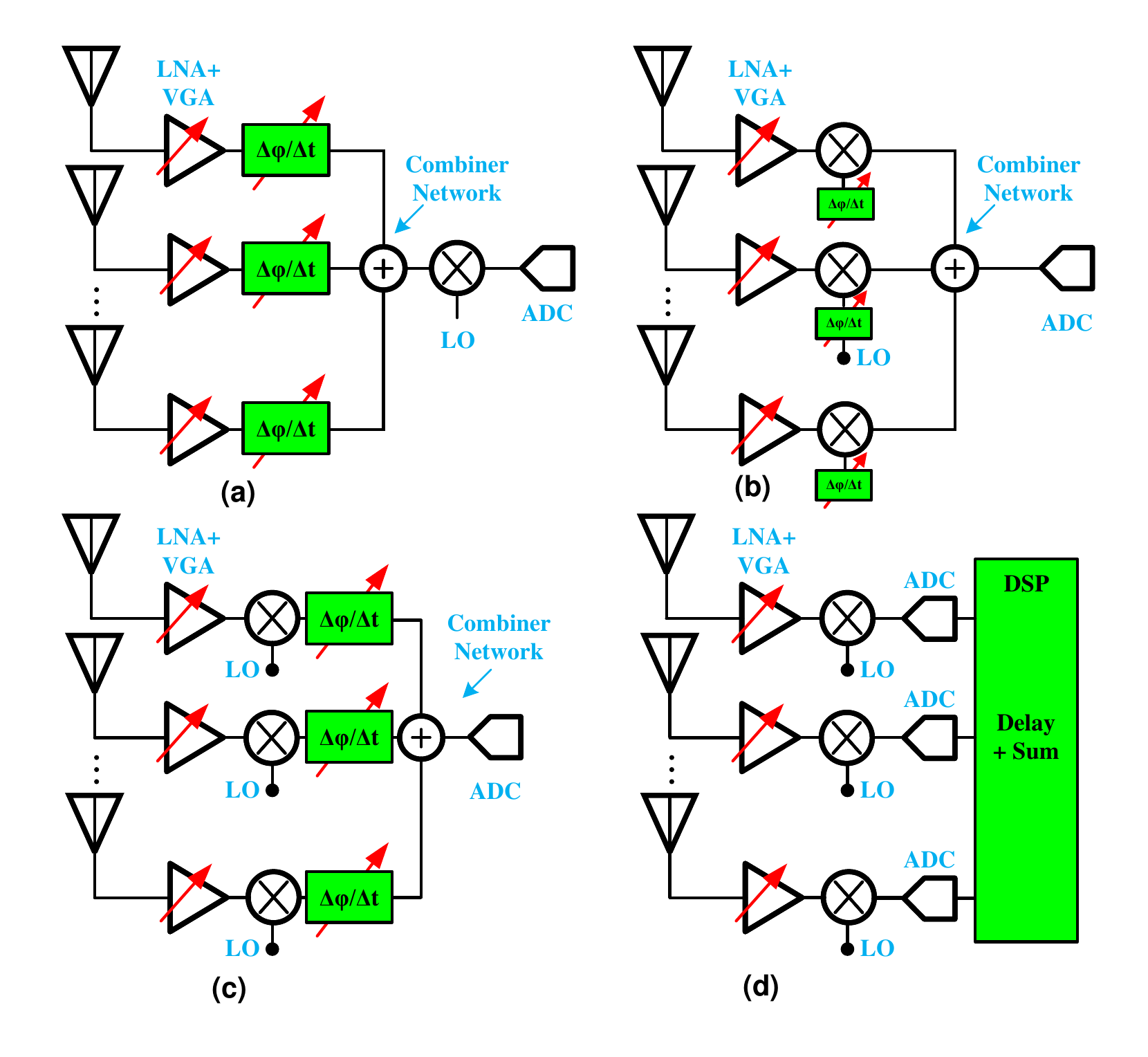}        
	\vspace{-8mm}
	\caption{Different receiver architectures for delay-sum beamforming (a) \ac{mmW}, (b) \ac{LO}, (c) Baseband, and (d) Digital domains.}
	\vspace{-3mm}
	\label{fig:arch_rx_dlysum}
\end{figure}

To address this fundamental issue in \ac{mmW} \ac{PAA}s, this article will introduce a fast beam training method leveraging \ac{TTD} arrays. Instead of suppressing the beam squint effect evident in \ac{mmW} arrays, we will exploit this effect to our advantage by introducing \textit{intentionally} large time delays in each antenna branch to realize frequency-dependent probing beams which can be exploited to accelerate the channel probing capability. These frequency-dependent beams will be fully controlled by adjusting the delay introduced in \ac{TTD} circuits \cite{Yan:TTD}. 

In summary, this article presents reconfigurable delay compensating circuits with large delay range and bandwidth, and finer delay resolution to realize wideband \ac{mmW} \ac{SSP}. Section~\ref{sec:syst:ttdarch} summarizes the \ac{TTD} \ac{SSP} architectures highlighting the challenges associated with realizing large delay compensation with fine resolution at different domains in the receiver chain. Section~\ref{sec:alg:beamtraining} discusses the fast beam-training algorithm and practical design considerations for analog \ac{PAA}s leveraging \ac{TTD} \ac{SSP} architecture. Section~\ref{sec:circuits} will present the hardware implementation of reconfigurable time delay units enabling different \ac{SSP} modes along with hardware validation methods for wideband \ac{PAA}s. Section~\ref{sec:futworks} presents future works expanding the proposed \ac{TTD}-based \ac{PAA}s for beam-training with planar arrays, multiple access applications, and standardization of wireless protocols. Section~\ref{sec:conclusions} concludes this article.


%
%

%
%

\section{Overview of \ac{mmW} \ac{TTD} \ac{PAA}s } 
\label{sec:syst:ttdarch} 


%

This section describes the possible architectural choices for \ac{TTD} \ac{PAA}s. The design of \ac{PAA}s are based on two principles: (1) phase/delay alignment of signals either in transmit or receive path, and (2) summation of phase/delay aligned signals. To align the received signals in a \ac{PAA}, delay elements should provide the proper delay to each received signal. The delay can be applied to the signal in \ac{mmW}, \ac{LO}, baseband, or digital domains. When the fractional bandwidth of the signal or the required delay range in the phased array system is small, the delay is approximated by a phase shift as discussed before. There are different topologies that are common in the implementation of \ac{PAA}  transceivers (Fig.~\ref{fig:arch_rx_dlysum}) such as \ac{RF} phase shifting \cite{sayginer2016, nafe2020,sadhu2017, Dunworth2018}, \ac{LO} phase shifting \cite{pang2019,wang2020}, \ac{BB} phase shifting \cite{mangravati2016,pellerano2019}, and digital phase shifting \cite{yang2018}. In the \ac{RF} phase-shifting architecture, after applying the required phase shift in each \ac{RF} path, the signals can be combined together. \ac{LO} phase shifting architecture uses the \ac{PS} in the \ac{LO} port of the mixer. In \ac{BB} and digital phase-shifting architecture, the received signals are aligned after downconversion, using analog circuitry in baseband or in a digital processor. Either of these architectures is a good candidate when the signal's fractional bandwidth is small, or the required delay range in the \ac{PAA} is small. 

\begin{table}
    \centering
	\caption{Survey of \ac{TTD} implementations at \ac{RF}/\ac{mmW}.}
	\vspace{0mm}
	\label{tab:surveyTTDRF}
	\vspace{-2mm}
	\includegraphics[width=1\columnwidth]{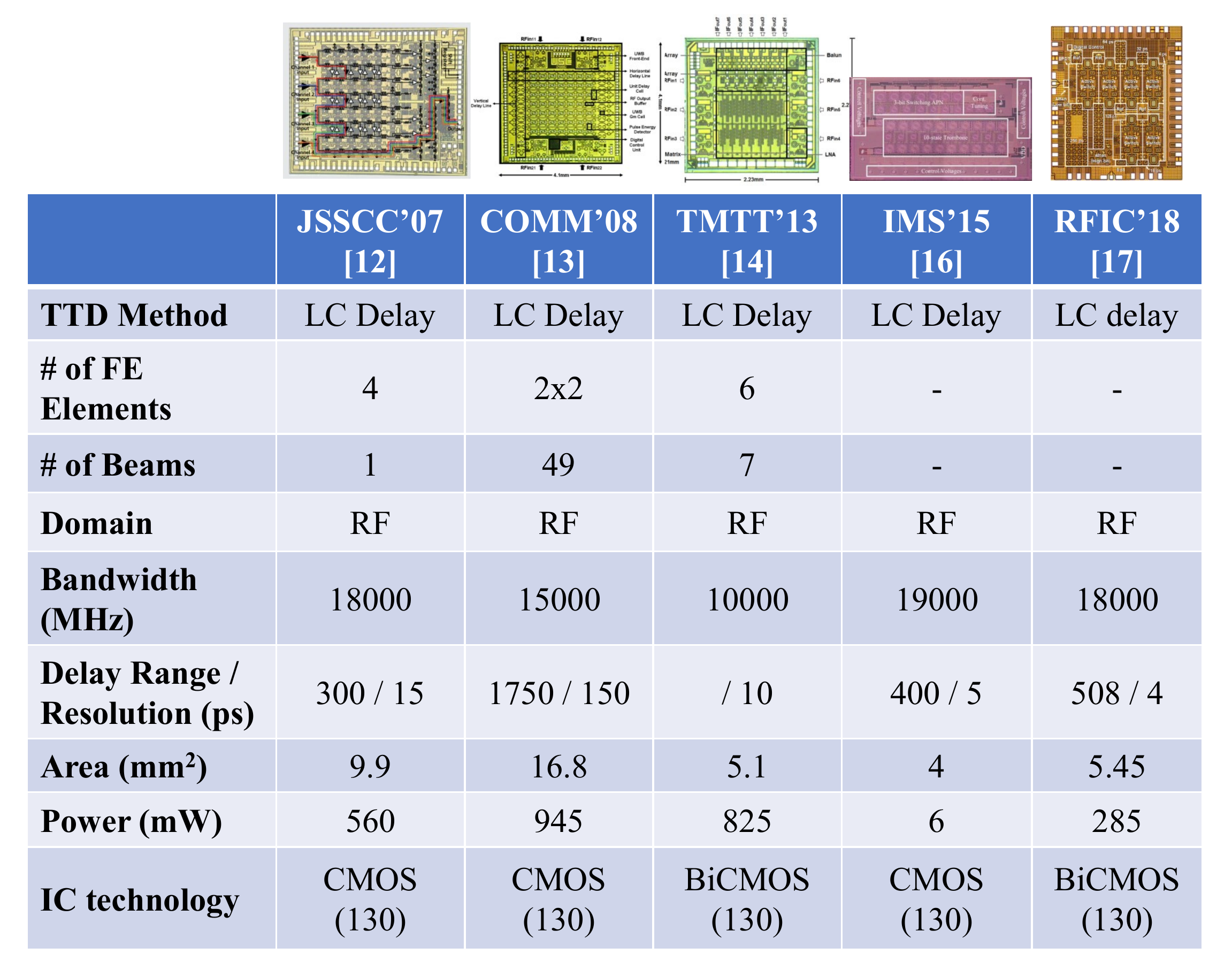}        
	\vspace{-6mm}
\end{table}

\begin{table*}
    \centering
	\vspace{-2mm}
	\caption{Survey of \ac{TTD} implementations at \ac{BB}.}
	\vspace{-3mm}
	\label{tab:surveyTTDBB}
	\includegraphics[width=1.6\columnwidth]{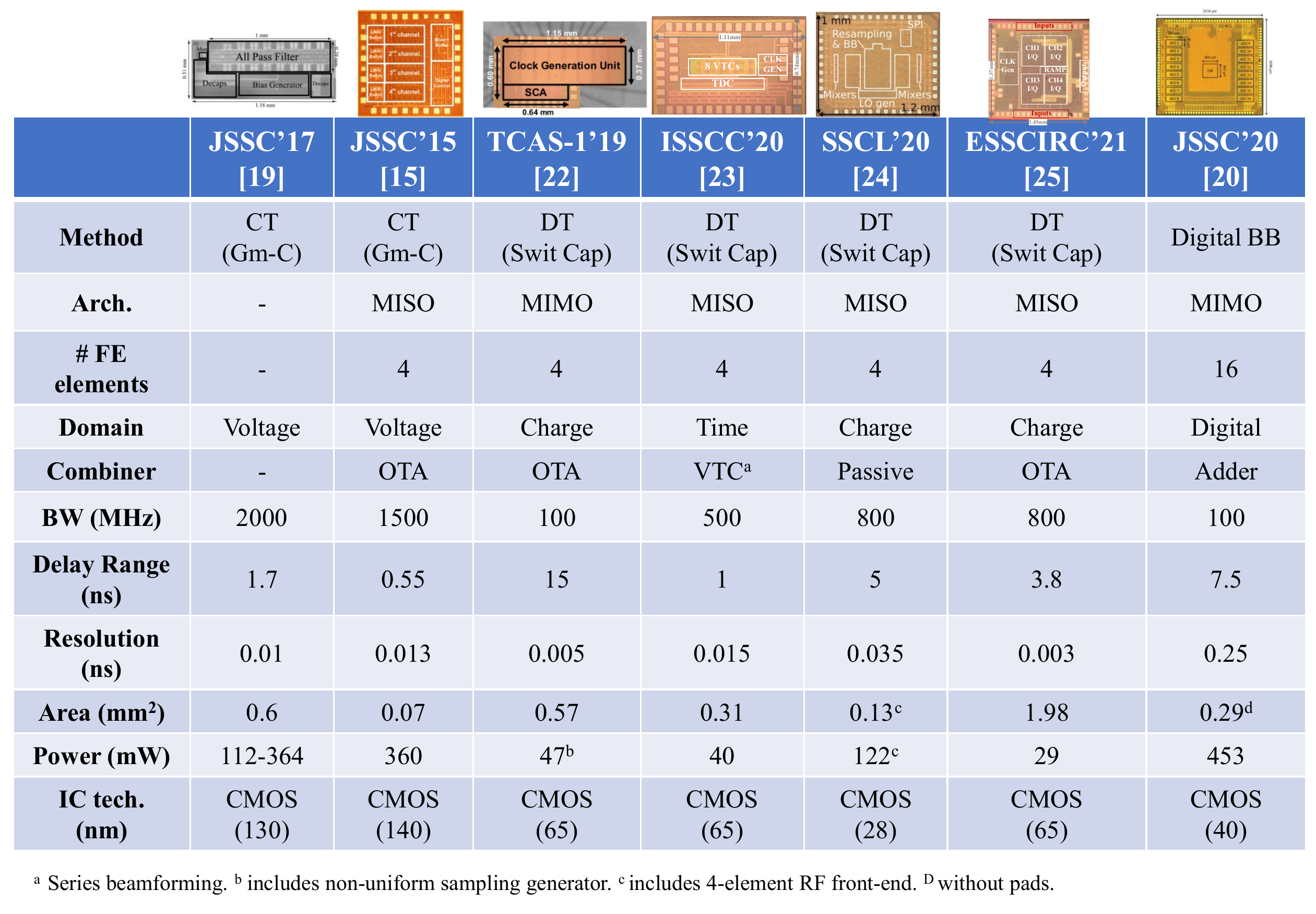}        
	\vspace{-2mm}
\end{table*}

For large delay-bandwidth product cases, delay units are necessary to prevent beam squint. Similar to a \ac{PS}-based \ac{PAA}s, the location of the time delay unit makes possible multiple architectures (\ac{mmW}/\ac{RF} path, \ac{LO} port of the mixer, \ac{BB}, or digital domains). In the \ac{mmW}/\ac{RF} \ac{TTD} architecture, the time delay unit is placed in the \ac{mmW}/\ac{RF} domain before the downconversion. The received signals at each receiver path can be combined constructively, as shown in Fig.~\ref{fig:arch_rx_dlysum}(a). The mixer is shared between multiple paths and only one \ac{LO} signal is necessary which allows significant area and power reduction. As the signals are combined before downconversion, interferers are removed due to spatial filtering, which reduces the linearity requirements of the blocks after the combiner. Although there are different implementations of the \ac{mmW}/\ac{RF} \ac{TTD}-based \ac{PAA}s, these implementations occupy a large area on the chip and there are serious limitations on the delay range. Table~\ref{tab:surveyTTDRF} highlights the integrated \ac{mmW}/\ac{RF} \ac{TTD} implementations over the past decade starting with the seminal work by Chu and Hashemi in 2007. Placing the \ac{TTD} element in the \ac{LO} path of the mixer does not resolve these limitations especially when handling wideband range. Digital \ac{TTD} \ac{PAA} is the same as a digital phased array and the challenges are the same such as high power consumption including the need for highly linear \ac{ADC} as well as other linear elements. However, there are promising \ac{TTD}-based \ac{PAA}s in \ac{BB} with large delay-bandwidth products \cite{ghaderi2019a,nagulu2021}. Table~\ref{tab:surveyTTDBB} highlights recent \ac{SOTA} \ac{TTD}-based \ac{PAA}s and delay lines implemented at \ac{BB}.  

\begin{figure}[t]
    \vspace{-2mm}
	\includegraphics[width=1\columnwidth]{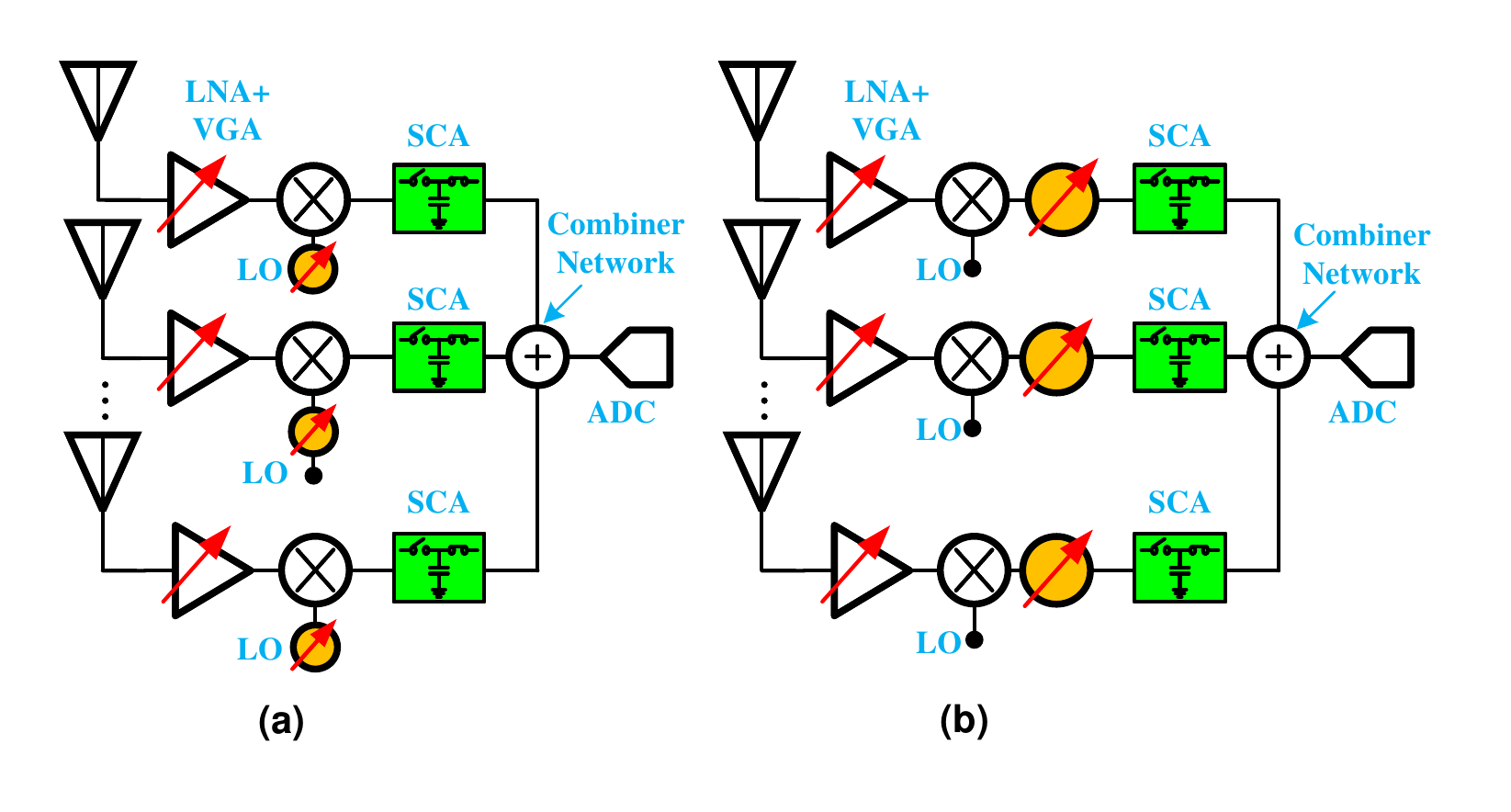}        
	\vspace{-8mm}
	\caption{\ac{BB} TTD with (a) \ac{LO}, and (b) \ac{BB} phase shifting.}
	\vspace{-4mm}
	\label{fig:arch_bbttd_lobb}
\end{figure}

%

In an array with \ac{BB} \ac{TTD} elements as shown in  Fig.~\ref{fig:arch_bbttd_lobb} \cite{ghaderi2019a}, instead of delaying the down-converted and phase shifted signals from the antennas followed by sampling and digitization, the signals are sampled at different time instants. Thus, the complexity of delaying signals is shifted to the clock path where precise and calibrated delays can be applied in the advanced semiconductor technology nodes. More importantly, a large delay range-to-resolution ratio can be easily realized while supporting a relatively larger number of antennas and bandwidth. The switched-capacitor adder based implementation requires multiple time-interleaved and delay compensated phases for formation of the beam \cite{ghaderi2019a}. In the sampling phase, the input signal from each channel is first sampled (with delayed time-interleaved clocks) on a sampling capacitor. After the last sampling phase, the stored charges on each capacitor corresponding to each channel (and each time-interleaved phase) are summed to form the beam. However, since time delay in \ac{BB} is not mathematically equivalent to the time delay in the \ac{RF} domain, a small phase shift is still necessary \cite{Jang2019 ,ghaderi2019a}. For example, if a time delay of $\tau_d$ is needed, it can be implemented in the \ac{BB} with a time delay equal to $\tau_d$ plus a phase shift equal to $-\omega_{\text{LO}}\tau_{d}$ \cite{ghaffari2014,ghaderi2019a,Jang2019}. The \ac{BB} \ac{TTD} architecture shown in Fig.~\ref{fig:arch_bbttd_lobb} is hereby referred as discrete-time \ac{TTD} \ac{SSP}.

A \ac{PS} can be implemented in the \ac{RF}, \ac{LO}, \ac{BB}, or digital domain to provide the required phase shift to the \ac{BB} time delay units. The \ac{mmW}/\ac{RF} \ac{PS} can be limited by the maximum delay-bandwidth product of the receiver. Digital implementation of the \ac{TTD} or \ac{PS} is similar to digital phase-shifting architecture with each \ac{RF} path requiring an \ac{ADC} which increases the power consumption of the whole system considerably. \ac{ADC}s should be linear enough to tolerate the large interferences which further increases the whole system's power consumption. Placing a \ac{PS} in \ac{BB} or \ac{LO} is another feasible option (Fig.~\ref{fig:arch_bbttd_lobb}). In these architectures, each \ac{RF} path has a dedicated mixer and an \ac{LO} signal. \ac{PS} in the \ac{LO} port of the mixer should operate at the \ac{LO} frequency. However, the \ac{PS} in the \ac{BB} or intermediate frequencies has to operate at lower frequencies. Though the loss and phase error of the \ac{PS} in \ac{RF} frequencies is higher than the \ac{PS} at \ac{BB}, the \ac{PS} at the \ac{BB} has to operate at a much higher fractional bandwidth while the \ac{PS} in \ac{LO} path only deals with a single tone signal (Fig.~\ref{fig:arch_bbttd_lobb}(a)). The \ac{PS} can be implemented in the \ac{LO} \cite{guan2004} as well as conventional \ac{PS} architectures. 

\ac{BB} \ac{PS} using vector summation is also another option. Vector summing \ac{PS} is based on the weighted summation of quadrature signals. Quadrature signal generation is possible using a quadrature coupler, polyphase filters, or using quadrature mixers. Quadrature couplers at the \ac{BB} occupies a large chip area and thus less preferred. Polyphase filters usually are narrow-band which contradicts with the large-delay bandwidth required. The use of quadrature downconversion mixers is another option though at the expense of routing complexities. 


The next section presents the framework of leveraging discrete-time \ac{TTD} \ac{SSP} for fast beam-training followed by details on hardware implementation and experimental validation of \ac{TTD} \ac{SSP}.

\section{Rainbow Beam-Training using Discrete-Time \ac{TTD} \ac{SSP}} 
\label{sec:alg:beamtraining}

This section describes the design of \ac{TTD} array parameters, which enable a training codebook of frequency-dependent beams to be synthesized. We also explain the corresponding frequency-domain \ac{DSP} algorithm \cite{Yan:TTD}. We further discuss the practical aspects of \ac{TTD} beam training including the sensitivity to hardware impairments, power allocation and the \ac{PAPR} problem in \ac{OFDM} systems, and achievable beam training distance.

\subsection{Introduction to Rainbow beam-training for \ac{mmW} \ac{PAA}s}
Recent work on \ac{mmW} and sub-THz beam-training has been focused on minimizing the required overhead using frequency-dependent beam steering/combining \cite{Ghasempour:lwa, Tan:tracking, hashemi2008a, Yan:TTD, Boljanovic:TTD, Boljanovic2021}. The main idea is to leverage different antenna architectures that can synthesize \textit{rainbow} beams to probe the entire angular range simultaneously. Each rainbow beam is associated with a different frequency, thus, information of the best steering/combining direction is embedded in the signal spectrum and it can be obtained through simple frequency-domain \ac{DSP}. In \cite{Ghasempour:lwa}, the authors proposed to use \ac{LWA} for a single-shot \ac{AoA} estimation.
In \cite{Tan:tracking}, a delay-phase array architecture was proposed for fast beam tracking with rainbow beams at THz frequencies.
\ac{TTD} array architectures in \cite{hashemi2008a} implemented delay elements in \ac{RF}, which are known to suffer from low scalability in terms of required area and power efficiency when the array size becomes large. In our recent work in \cite{Yan:TTD, Boljanovic:TTD, Boljanovic2021}, we proposed a scalable \ac{TTD} arrays with delay elements implemented in baseband for a single-symbol \ac{mmW} beam training in an \ac{OFDM} system. The key idea was to exploit signal delaying in each antenna branch to exacerbate the wideband spatial effect (beam squint) among subcarriers in a controllable manner, i.e., to create a frequency-dependent codebook which probes all angular directions at once. This is achieved by properly configuring the delay and phase taps in a \ac{TTD} array. 

\begin{figure}
    \vspace{-2mm}
    \begin{center}
        \includegraphics[width=0.45\textwidth]{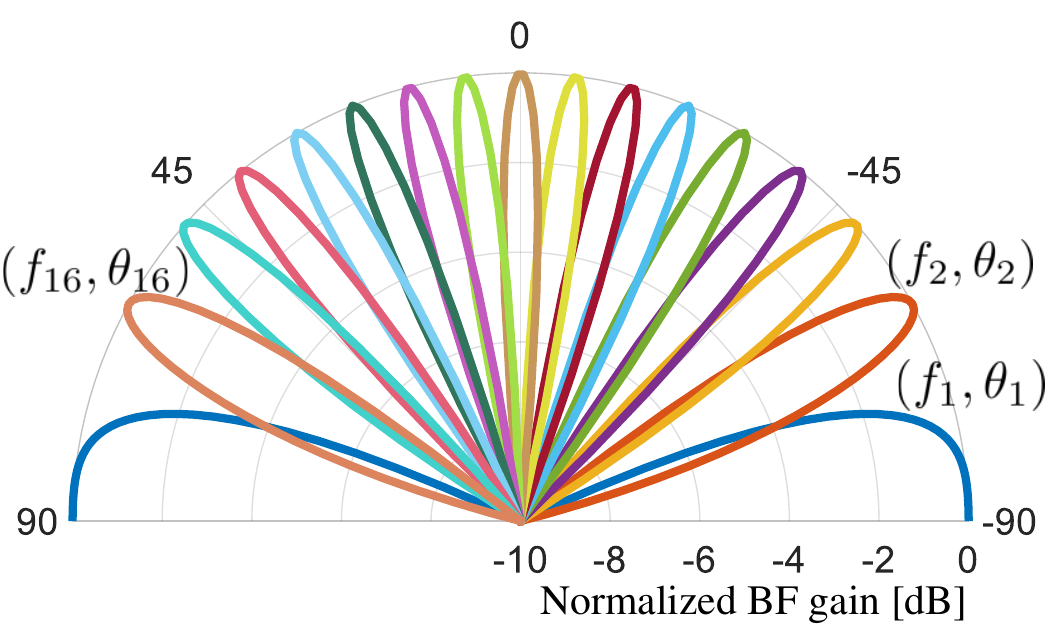}
    \end{center}
    \vspace{-2mm}
    \caption{Resulting \ac{TTD} codebook, assuming $N_{\R}=16$, $D=16$, $M=16$. Subcarriers $f_m,~\forall m$, are associated with their corresponding angles $\theta_m,~\forall m$.}
    \vspace{-4mm}
    \label{fig:codebook}
\end{figure}

\subsection{\ac{TTD} codebook and \ac{DSP} algorithm design}
\label{sec:codebook_and_dsp}

In our recent work we introduced \ac{TTD}-based beam-training at the \ac{UE} side that requires only one \ac{OFDM} symbol \cite{Yan:TTD, Boljanovic:TTD, Boljanovic2021}. We assume that the \ac{BS} with $N_{\T}$ antennas uses a fixed frequency-flat precoder $\mathbf{v}\in \mathbb{C}^{N_{\T}}$ designed according to the previously estimated \ac{AoD}. The received signal $Y[m]$ at the $m$-th \ac{OFDM} subcarrier can be expressed as follows
\begin{equation}
    Y[m] = \mathbf{w}^{\hermitian}[m] \mathbf{H}[m] \mathbf{v}s[m] + \mathbf{w}^{\hermitian}[m]\mathbf{n}[m],
    \label{eq:received_signal}
\end{equation}
where $\mathbf{w}[m]\in\mathbb{C}^{N_{\R}}$ is a frequency-dependent \ac{UE} \ac{TTD} combiner, $\mathbf{H}[m]\in\mathbb{C}^{N_{\R}\times N_{\T}}$ is a channel matrix, $s[m]$ is a pilot, and $\mathbf{n}[m]\sim \mathcal{CN}(0,\sigmaN^2\mathbf{I}_{\R})$ is a noise vector at the $m$-th subcarrier, respectively. The $n$-th element of $\mathbf{w}$ is defined as
\begin{equation}
    \left[\mathbf{w}[m]\right]_n = \alpha_n\exp\left(-j\left(2\pi (f_m-\fc)\tau_n + \phi_n \right)\right).
    \label{eq:combiner}
\end{equation}
where $\alpha_n=1$, $\tau_n$, and $\phi_n$ are the gain, delay tap, and phase tap in the $n$-th antenna branch, respectively.

Assuming the bandwith $\BW$ and a \ac{ULA} with a frequency-flat spatial response at the \ac{UE} side, it was shown in \cite{Yan:TTD} that a codebook of rainbow beams which cover the entire angular range $[-\pi/2, \pi/2]$ can be created by setting the delay taps $\tau_n,~\forall n$, as follows
\begin{align}
    \tau_n = (n-1)\Delta\tau,~\forall n,~~\text{where}~~\Delta \tau=1/\BW,
    \label{eq:delay_taps}
\end{align}
Since the taps in (\ref{eq:delay_taps}) are proportional to the Nyquist sampling period, the codebook can be obtained without the need to implement a fractional \ac{ADC} sampling. In such a codebook, the rainbow beam associated with the subcarrier frequency $f_m$ is pointing in the angle $\theta_m$ given as follows \cite{Yan:TTD}
\begin{equation}
    \theta_m = \sin^{-1}\left(\mathrm{mod}(2f_m \Delta\tau + 1, 2) - 1 \right).
    \label{eq:ang_freq_mapping}
\end{equation}
An example of the resulting \ac{TTD} codebook for a \ac{UE} with $N_{\R}=16$ antenna elements is provided in Fig.~\ref{fig:codebook}. To probe $D=N_{\R}=16$ directions, $M=16$ subcarriers need to be used. The codebook can be designed to be more robust in frequency-selective channels by mapping $R$ different subcarriers to each probed direction. This is achieved by increasing the delay difference between antenna elements $R$ times, i.e., $\Delta\tau=R/\BW$ \cite{Boljanovic:TTD}. With $R$ being the codebook diversity order and $D$ the number of probed directions, \ac{TTD} beam training requires a waveform with $M=DR$ loaded subcarriers.

\begin{table}[t]
\vspace{0mm}
\centering
\caption{\small Assumed parameters for numerical evaluation.} 
\label{tab:sim_par}
\vspace{-2mm}
{   
    \footnotesize { 
    \begin{tabular}{|c|c|c|} 
    \hline
        Symbol & Value & Description \\
        \hline
        \hline
        $\fc$  &  60~GHz & Carrier frequency\\
        \hline
        $\BW$  & 2~GHz & Bandwidth  \\
        \hline 
        $M_{\tot}$ &  4096  & Total number of subcarriers\\
        \hline
        $N_{\T}$ &  128  & Number of \ac{BS} antennas\\
        \hline
        $N_{\R}$ &  16  & Number of \ac{UE} antennas\\
        \hline
        $D$ &  32  & Number of probed directions\\
        \hline
        $R$ &  4  &  Codebook diversity order\\
        \hline
        $M$ ($=DR$) &  128  &  Number of used subcarriers\\
        \hline
        $Q$ &  1024  & Dictionary size in \cite{Boljanovic:TTD}\\
        \hline
        \ac{SNR}  &  0~dB  &  Signal-to-noise ratio\\
        \hline
    \end{tabular} \\
}
}
\vspace{-4mm}
\end{table}

The pointing angles $\theta_m,~\forall m$, of rainbow beams can be \textit{jointly} rotated by using the frequency-flat \ac{PS}s in the \ac{TTD} array. A rotation of $\theta_{\rot}$ requires the phase taps $\phi_n,~\forall n$, to be set as follows
\begin{equation}
    \phi_n = (n-1)\Delta \phi,~\forall n,~~\text{where}~~\Delta\phi = \pi \sin(\theta_{\rot}).
    \label{eq:phase_taps}
\end{equation}
The rotated angles can then be expressed as $\theta_m^{\prime} = \theta_m + \theta_{\rot},~\forall m$.

With this codebook and described frequency-to-angle mapping, the information of the dominant propagation angle can be acquired with simple frequency-domain \ac{DSP}. Given the received signal in (\ref{eq:received_signal}), a coarse \ac{AoA} estimate $\hat{\theta}$ is given as $\hat{\theta}=\theta_{m^*}^{\prime}$, where the angle $\theta_{m^*}^{\prime}$ corresponds to the subcarrier with the strongest received signal power. Mathematically, the estimation can be expressed as
\begin{equation}
    \hat{\theta}=\theta_{m^*}^{\prime},~~\text{where}~~m^* = \argmax_{m} \left|Y[m]\right|^2.
    \label{eq:angle_est}
\end{equation}
If the codebook diversity is $R$, the power $\left|Y[m]\right|^2$ is averaged across all $R$ subcarriers $m$ mapped into the same direction before angle estimation.

The estimation accuracy can be improved by designing a high-resolution dictionary-based algorithm that relies on the codebook with diversity order $R$ \cite{Boljanovic:TTD}. The average received signal power in all probed directions can be correlated with an oversampled dictionary of \ac{UE} beamforming gains to obtain an accurate angle estimate.

\subsection{Practical considerations in \ac{TTD} beam-training}

\subsubsection{Sensitivity to hardware impairments}

In practice, errors in different hardware components can distort the delay taps in (\ref{eq:delay_taps}) and phase taps in (\ref{eq:phase_taps}) and thus affect the designed beam training codebook. We assume that delay errors can be modeled as Gaussian random variables, such that the distorted taps are $\tilde{\tau}_n\sim \mathcal{N}(\tau_n, \sigma_{\T}^2)$, where $\sigma_{\T}^2$ is the error variance. The phase errors occur due to imperfect \ac{PS}s, LOs, imbalance between in-phase and quadrature-phase samples, or other hardware errors. Similarly as with the delays, we model the distorted phase taps as $\tilde{\phi}_n\sim \mathcal{N}(\phi_n, \sigma_{\text{P}}^2)$, where $\sigma_{\text{P}}^2$ is the error variance. In addition to the phase and delay errors, we consider distorted gain $\tilde{\alpha}_n$ in all antennas branches. The gains are modeled with a log-normal distribution, i.e., $10\log_{10}(\tilde{\alpha}_n)\sim\mathcal{N}\left(0,\sigma_{\text{A}}^2\right),~\forall n$.

\begin{figure}
    \begin{center}
        \includegraphics[width=0.485\textwidth]{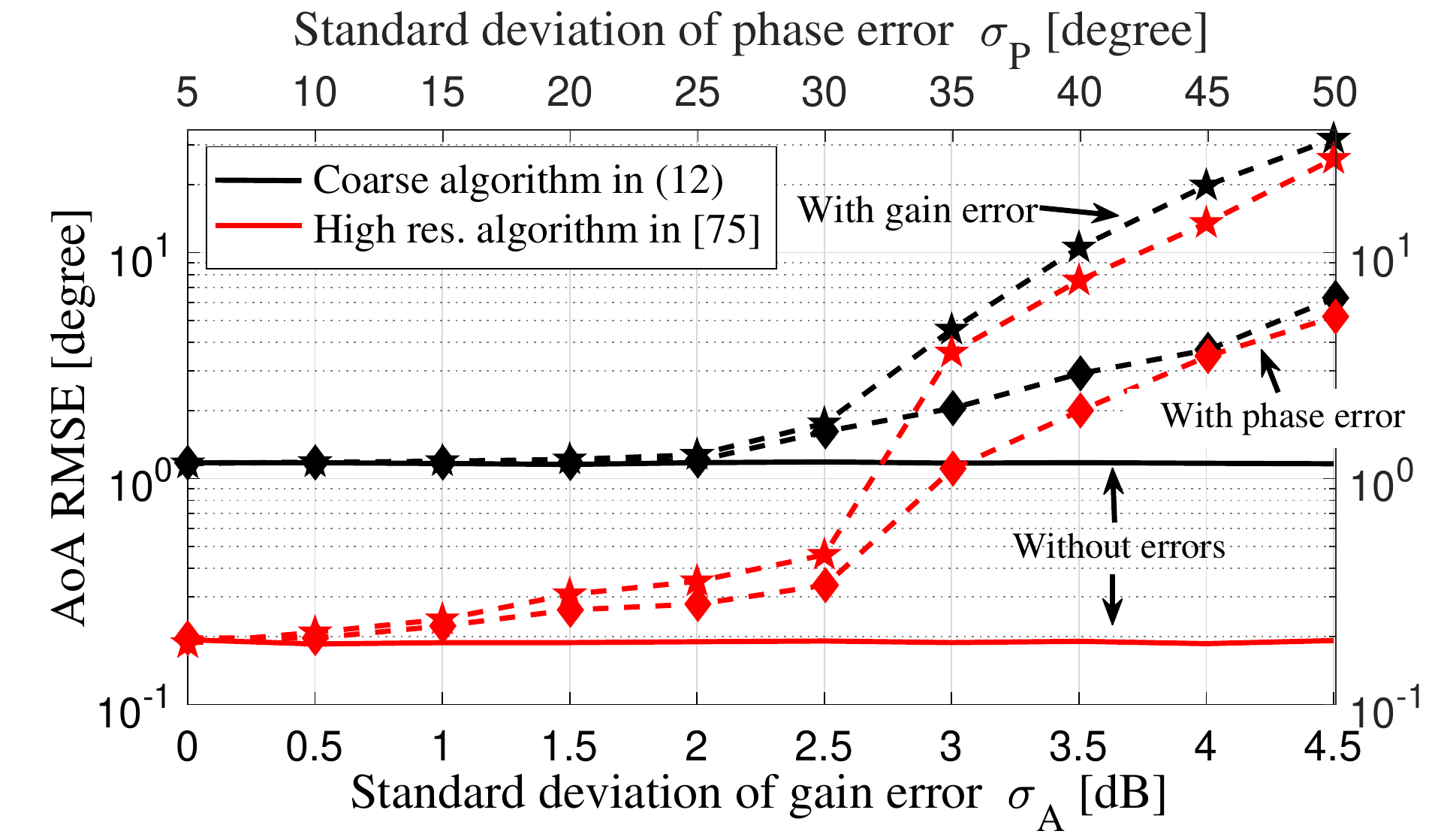}
    \end{center}
    \vspace{-3mm}
    \caption{Impact of gain and phase errors on performance of two \ac{TTD} beam training algorithms \cite{Boljanovic:TTD}. The curves with gain error (dashed with stars) and phase error (dashed with diamonds) are associated with the lower and upper x-axis, respectively.}
    \vspace{-2mm}
    \label{fig:gain_phase}
\end{figure}

\begin{figure}[t]
    \begin{center}
        \includegraphics[width=0.485\textwidth]{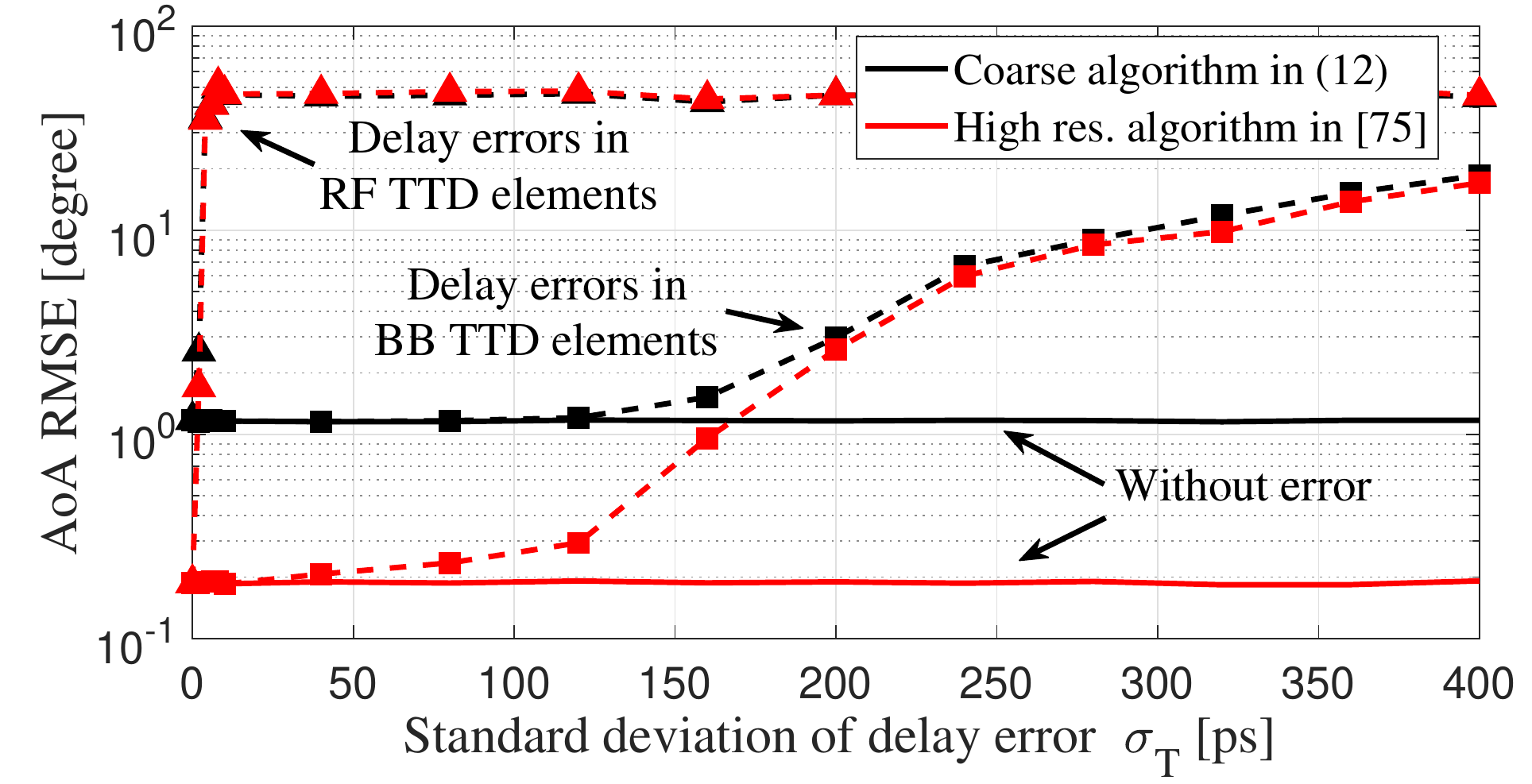}
    \end{center}
    \vspace{-4mm}
    \caption{Impact of delay error on performance of two \ac{TTD} beam training algorithms \cite{Boljanovic:TTD}.}
    \vspace{-4mm}
    \label{fig:delay}
\end{figure}

We have numerically evaluated the impact of all three hardware errors on the \ac{TTD} beam training performance. Specifically, we compared the impact on two algorithms, including the coarse angle estimation algorithm introduced in the previous subsection and high-resolution algorithm in \cite{Boljanovic:TTD}. The assumed simulation parameters are summarized in Table~\ref{tab:sim_par}. In Fig.~\ref{fig:gain_phase}, we study the impact of gain and phase errors on the \ac{RMSE} of \ac{AoA} estimation, under no delay error. For both algorithms, severe performance degradation occurs when the standard deviation of the gain error $\sigma_{\text{A}}\geq 2.5$ dB and phase error $\sigma_{\text{P}}\geq 30^{\circ}$. In Fig.~\ref{fig:delay}, the impact of \ac{TTD} delay error is presented. With the proposed TDD architecture where time delay units are implemented in baseband, both algorithms are robust to delay errors with standard deviation of up to $\sigma_{\T}=$ \SI{125}{\pico\second}. In comparison, both algorithms start to show severe degradation with only $\sigma_{\T}=$\SI{1.5}{\pico\second} in TDD architectures with \ac{RF} time delay units studied in previous work \cite{Boljanovic:TTD}. This result indicates that both algorithms have more relaxed specifications for baseband implementation of \ac{TTD} units compared to the \ac{RF} \ac{TTD} arrays. 

\subsubsection{Power allocation and supported distances}

\begin{figure}[t]
    \begin{center}
        \includegraphics[width=0.485\textwidth]{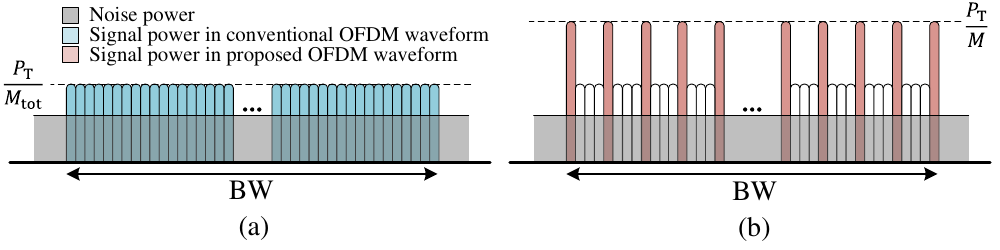}
    \end{center}
    \vspace{-4mm}
    \caption{Illustration of \ac{OFDM} power allocation in (a) conventional beam training and (b) \ac{TTD} beam training.}
    \vspace{0mm}
    \label{fig:power_allocation}
\end{figure}

\begin{figure}
    \begin{center}
        \includegraphics[width=0.46\textwidth]{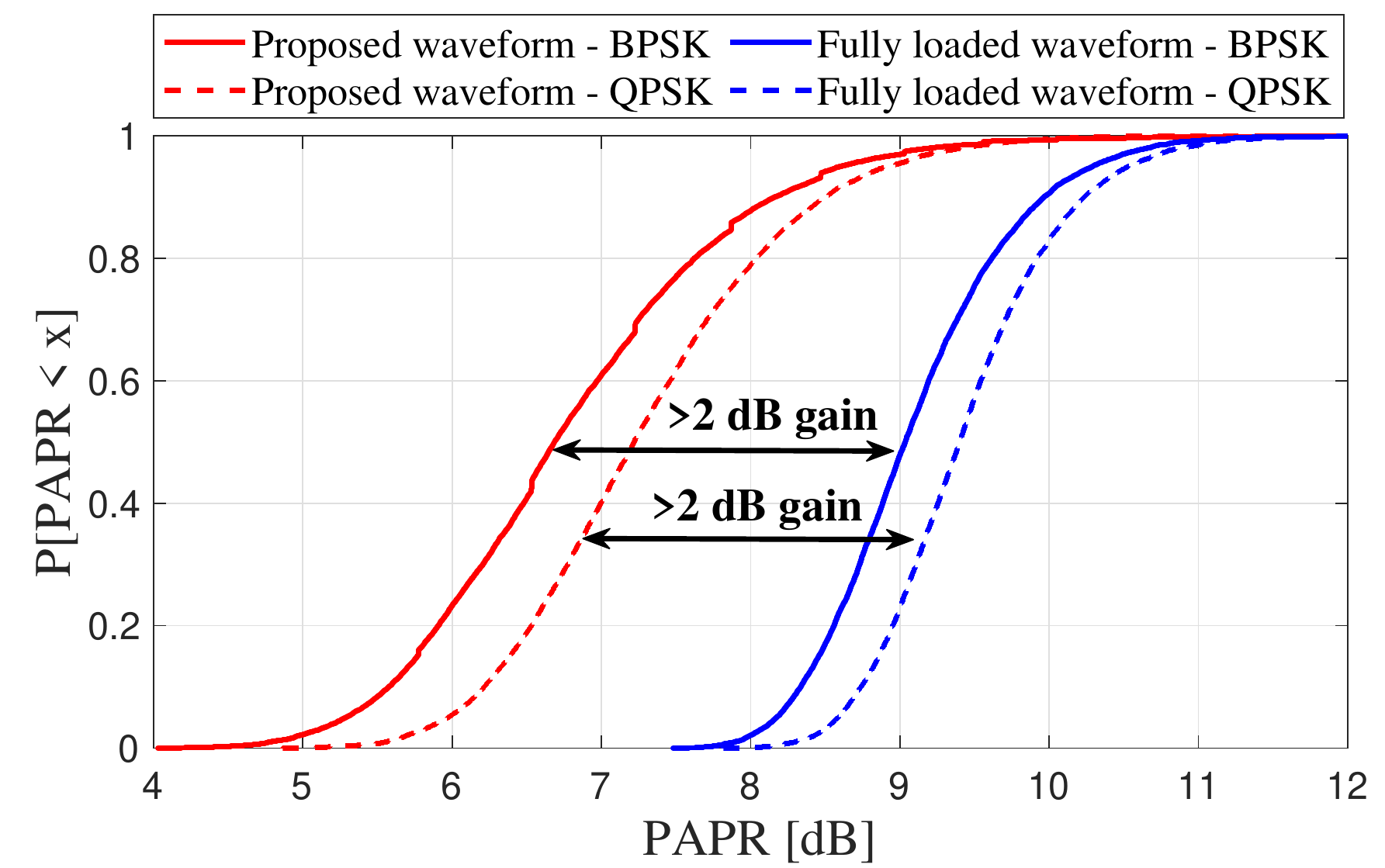}
    \end{center}
    \vspace{-4mm}
    \caption{\ac{PAPR} comparison between proposed and fully loaded \ac{OFDM} waveforms, assuming \ac{BPSK} or \ac{QPSK} symbols.}
    \vspace{-4mm}
    \label{fig:papr_cdf}
\end{figure}

The proposed frequency-dependent beam-training uses an \ac{OFDM} based waveform where only a subset of $M$ out of $M_{\tot}$ ($M < M_{\tot}$) subcarriers is loaded at the \ac{BS}. This allows the transmit signal power to be allocated to a lower number of subcarriers, which can increase the \ac{SNR} per subcarrier compared to that in conventional beam-training, as illustrated in Fig.~\ref{fig:power_allocation}. Specifically, when the total transmit power $P_{\T}$ is divided among $M$ subcarriers, the \ac{SNR} per subcarrier is given by the following expression
\begin{equation}
    \text{SNR}_{\text{sc}} = \frac{G_{\T}G_{\R}\lambda^2}{(4\pi d)^2}\frac{P_{\T}}{ \Delta \BW N_0 M},
    \label{eq:snr_per_sc}
\end{equation}
assuming a free-space path loss model. The terms $G_{\T}=20\log_{10}\left(N_{\T} \right)$, $G_{\R}=20\log_{10}\left(N_{\R} \right)$, $\lambda$, and $d$ represent the transmit beamforming gain, receive beamforming gain, wavelength, and distance between the \ac{BS} and \ac{UE}. Power spectral density of the noise is denoted as $N_0$, while $\Delta \BW = \BW/(M_{\tot}-1)$ represents the subcarrier spacing. Note that the \ac{SNR} per subcarrier is $M_{\tot}/M$ times larger than with a fully loaded \ac{OFDM} waveform. Since the proposed \ac{DSP} algorithm considers only $M$ loaded subcarriers with high \ac{SNR}, angle estimation in (\ref{eq:angle_est}) is not noise-limited. Further, due to a lower number of used subcarriers, the proposed \ac{OFDM} waveform for \ac{TTD} beam training results in a more than $2$dB lower \ac{PAPR} than a fully loaded \ac{OFDM} waveform, as presented in Fig.~\ref{fig:papr_cdf}, where we assumed the same simulation parameters as in the previous subsection. We used a cyclic prefix of 128 samples and assumed that subcarriers are loaded either with \ac{BPSK} or \ac{QPSK} symbols.

\begin{figure}[t]
    \begin{center}
        \includegraphics[width=0.485\textwidth]{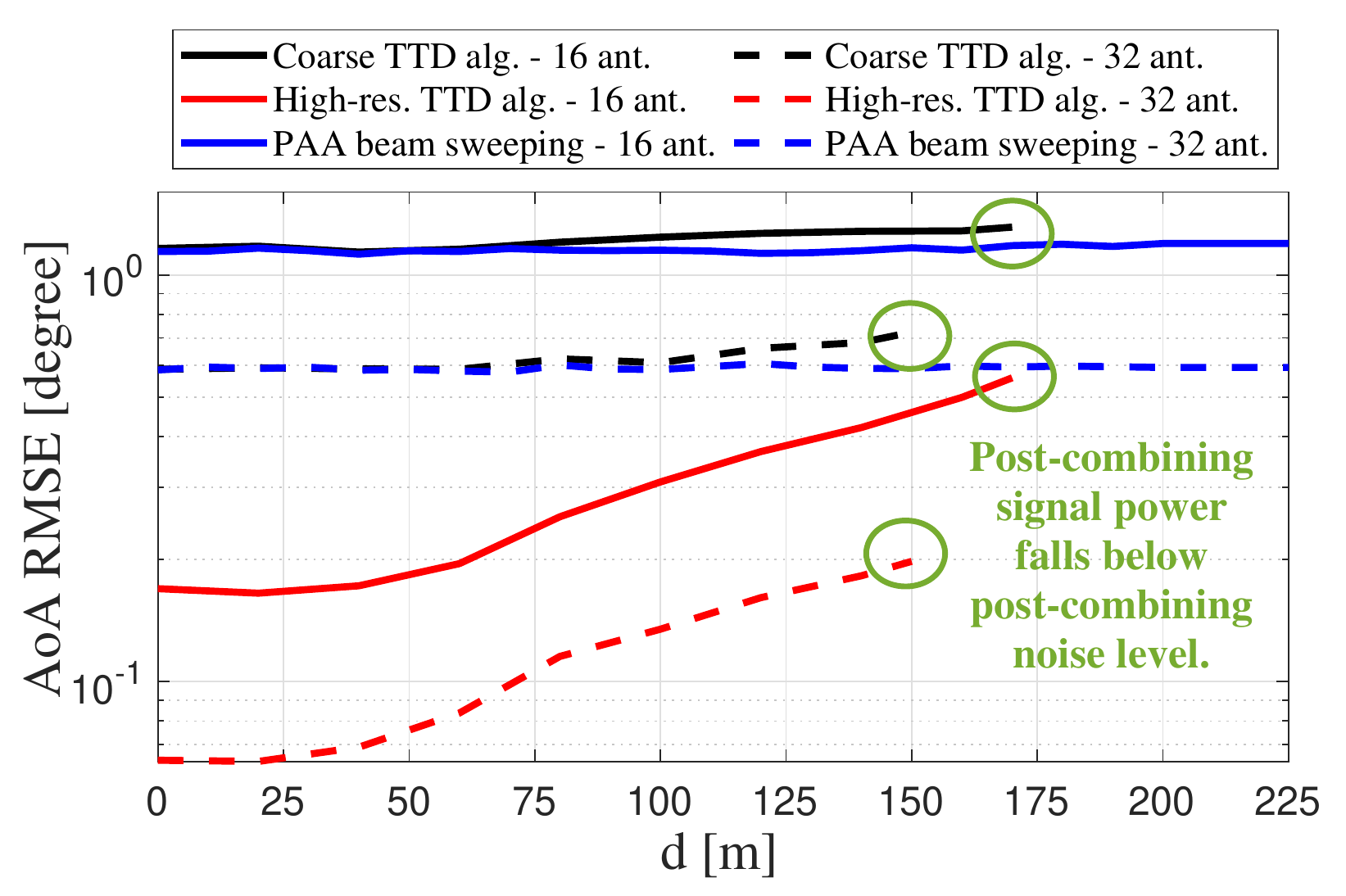}
    \end{center}
    \vspace{-4mm}
    \caption{Comparison between \ac{TTD} algorithms and \ac{PAA}-based beam sweeping in terms of \ac{AoA} RMSE at different distances between the \ac{BS} and \ac{UE}.}
    \vspace{-4mm}
    \label{fig:rmse_vs_distance}
\end{figure}

Although the proposed frequency-dependent beam-training can benefit from a higher \ac{SNR} per subcarrier, the supported distances between the \ac{BS} and \ac{UE} are reduced compared to the \ac{PAA} beam sweeping approach. This is because the \ac{UE} has a lower \textit{total} received signal power. Specifically, with the codebook in Fig.~\ref{fig:codebook}, which probes $D$ different directions, roughly $1/D$ of the signal power is received per direction. The received signal power corresponds to the subcarriers that probe the dominant \ac{AoA}. To evaluate the impact of lower received signal power on the beam-training performance, we compared the \ac{AoA} \ac{RMSE} of \ac{TTD}-based algorithms and \ac{PAA}-based beam sweeping at different distances between the \ac{BS} and \ac{UE}. We assumed a non-line-of-sight scenario and modeled the path loss according to mmMAGIC channel \cite{5G:mmMagic}. We considered two array sizes at the \ac{UE}, including $N_{\R}=16$ and $N_{\R}=32$. With $N_{\R}=16$, $D=32$ directions are probed, while with $N_{\R}=32$, $D=64$ directions are probed. Other parameters are the same as in the previous subsection. The comparison results are presented in Fig.~\ref{fig:rmse_vs_distance}. \ac{PAA}-based beam sweeping combines the entire signal bandwidth in each probed direction and thus achieves a high received signal power and reliable angle estimation at different distances. Nevertheless, it requires large overheads of $32$ and $64$ \ac{OFDM} symbols to do so when $N_{\R}=16$ and $N_{\R}=32$, respectively. On the other hand, \ac{TTD} beam training algorithms are based on a single \ac{OFDM} symbol and their performance is comparable to that of exhaustive \ac{PAA} beam sweeping. However, supported distances in \ac{TTD} beam-training are reduced, assuming that the signal cannot be detected when its total post-combining power falls below post-combining noise level. In our simulations, this happens at $d=170$m for $N_{\R}=16$ and $d=150$m for $N_{\R}=32$, as highlighted in Fig.~\ref{fig:rmse_vs_distance}. We note that with a more sophisticated detection algorithm that exploits the waveform structure, beam training capabilities and supported distances of \ac{TTD} arrays could increase. We leave a more detailed theoretical study of supported distances and better detection algorithms for future work.




%
%

\section{Reconfigurable Discrete-time \ac{TTD} \ac{SSP} Hardware Design Considerations}
\label{sec:circuits}

In this section, we describe the design and hardware implementation of reconfigurable time delay units in the discrete-time \ac{TTD} \ac{SSP} shown in Fig.~\ref{fig:arch_bbttd_lobb} applied to both the data communications as well as beam-training \ac{SSP} modes.

\subsection{Important sub-systems in discrete-time \ac{TTD} \ac{SSP}}

\begin{figure}
    \begin{center}
        \includegraphics[width=0.45\textwidth]{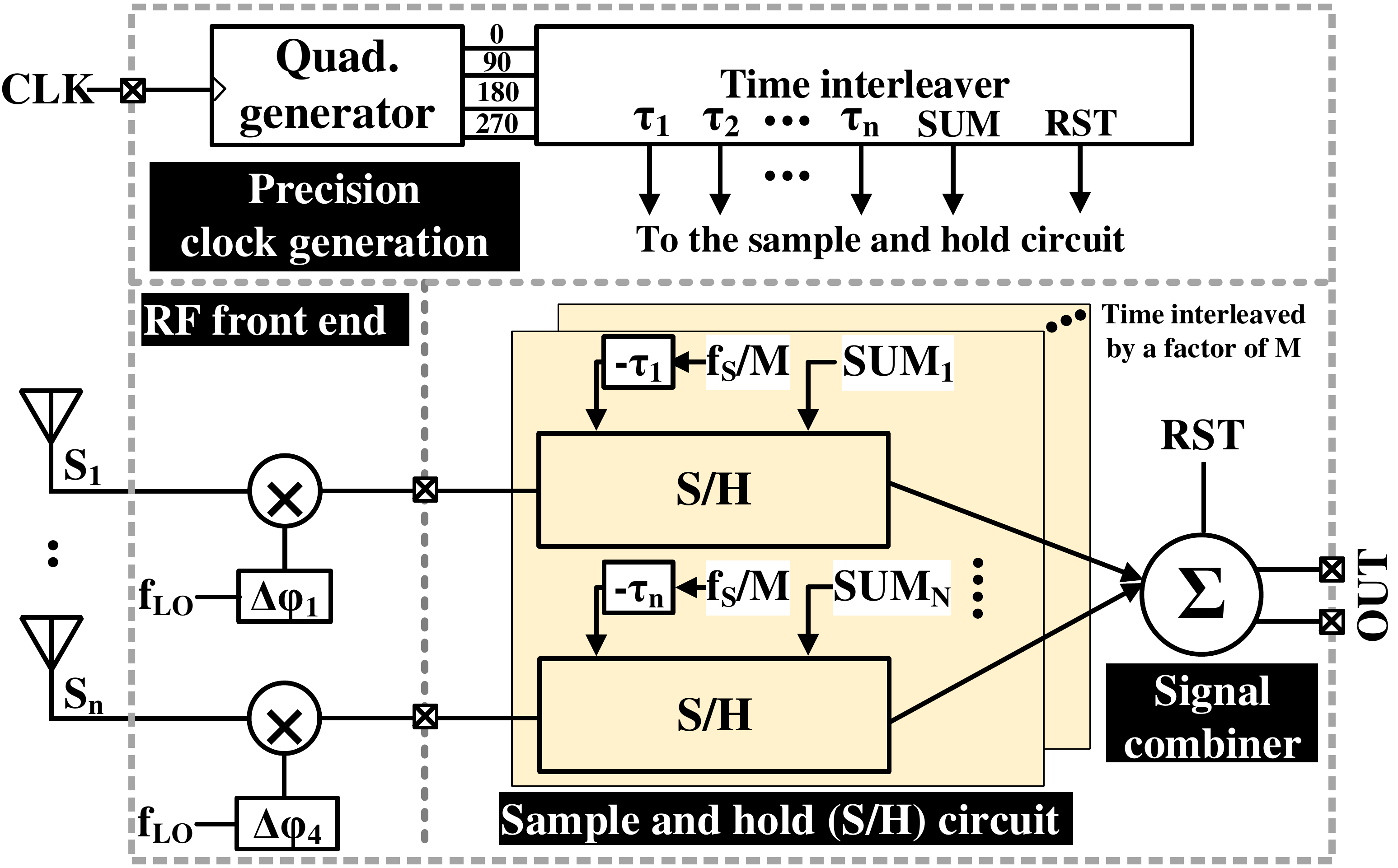}
    \end{center}
    \vspace{-3mm}
    \caption{Generic \ac{TTD} \ac{SSP} architecture.}
    \vspace{-4mm}
    \label{fig:ttdssparch}
\end{figure}

The analog array though more energy efficient when compared to hybrid \ac{TTD} and digital arrays \cite{Boljanovic2021} require larger unity-gain bandwidth amplifiers to have similar performance. The circuit performance is further constrained with the parasitic capacitance, routing losses, crosstalk, and any possible mismatches during fabrication evident at multiple levels including silicon, packaging, or printed circuit board. Design considerations for the key blocks are described here to ensure low-power consumption is upheld for analog arrays as compared to hybrid or digital \ac{TTD} \ac{SSP} \cite{Boljanovic2021} including its ability to achieve fractional delays with high precision. This section presents design considerations for the key design blocks: (i) sample-and-hold, (ii) wideband signal combiners, and (iii) precision clock generation. Fig.~\ref{fig:ttdssparch} shows a more detailed illustration of the discrete-time \ac{TTD} \ac{SSP}. Interested readers can refer to \cite{Boljanovic2021, Lin2021, ghaderi2019a, ghaderi2019b} for further understanding.  

\subsubsection{Sample-and-hold} 
The design constraints for the input sample-and-hold draws parallel to the design requirements of a high-speed time-interleaved \ac{ADC} \cite{kull2018}. Different interleaver configurations can be analyzed based on a simplified switch model in which the switch resistance and capacitance are linked to the technology. To reduce effect of sampling jitter, it is preferable to sample first followed by subsampling however at the cost of additional source followers which can significantly impact power consumption and linearity. To relax the power consumption, the first sampler can be interleaved (typically by 2 or 4) at the cost of slightly higher mismatch and jitter \cite{greshishchev2010}.

\begin{figure}[t]
\vspace{-4mm}
\centering
\begin{tabular}{cc}
\vspace{-2mm}
\subfloat[]{%
  \includegraphics[clip,width=0.8\columnwidth]{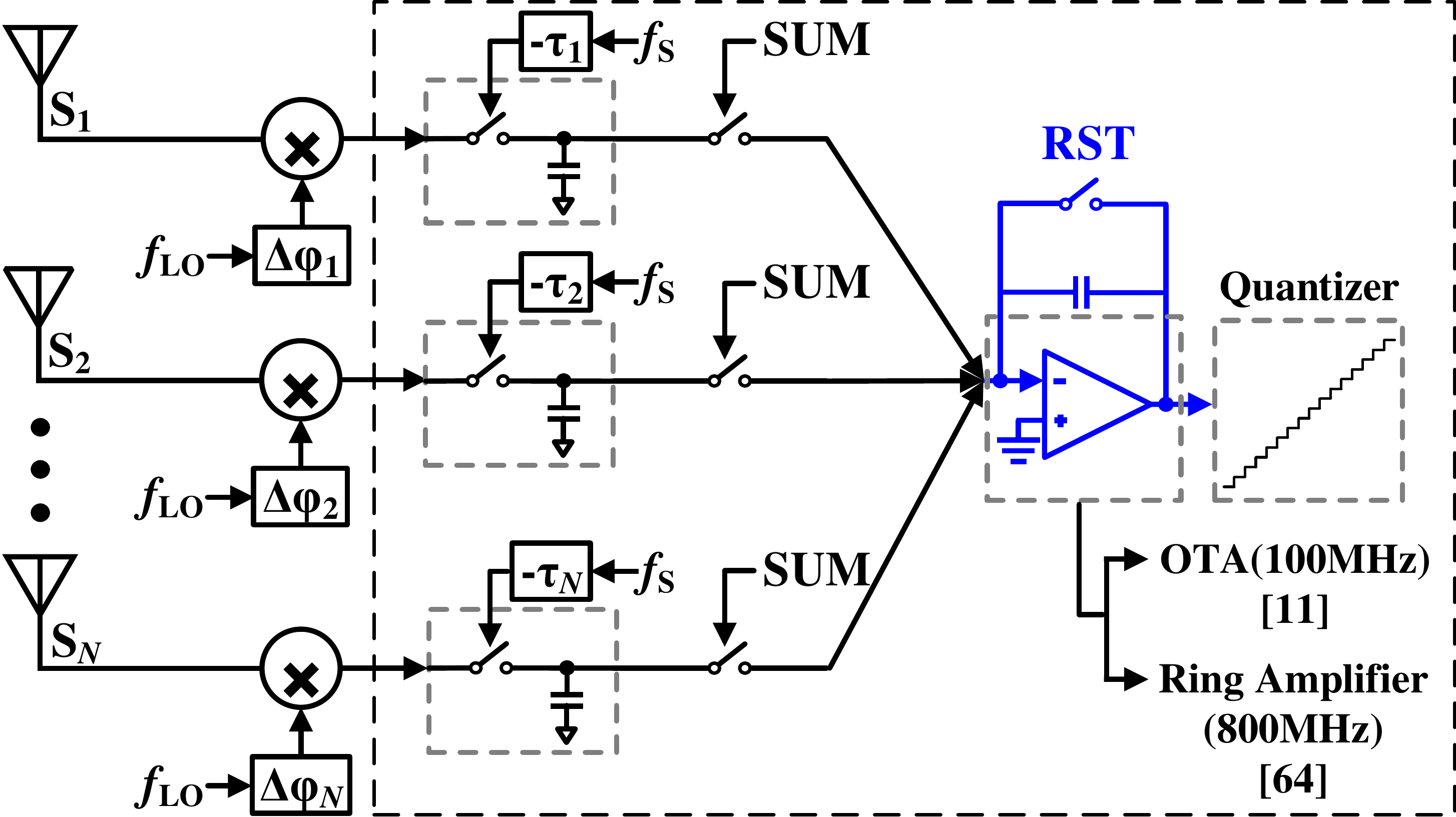}%
}\\
\vspace{-4mm}
\subfloat[]{%
  \includegraphics[clip,width=0.8\columnwidth]{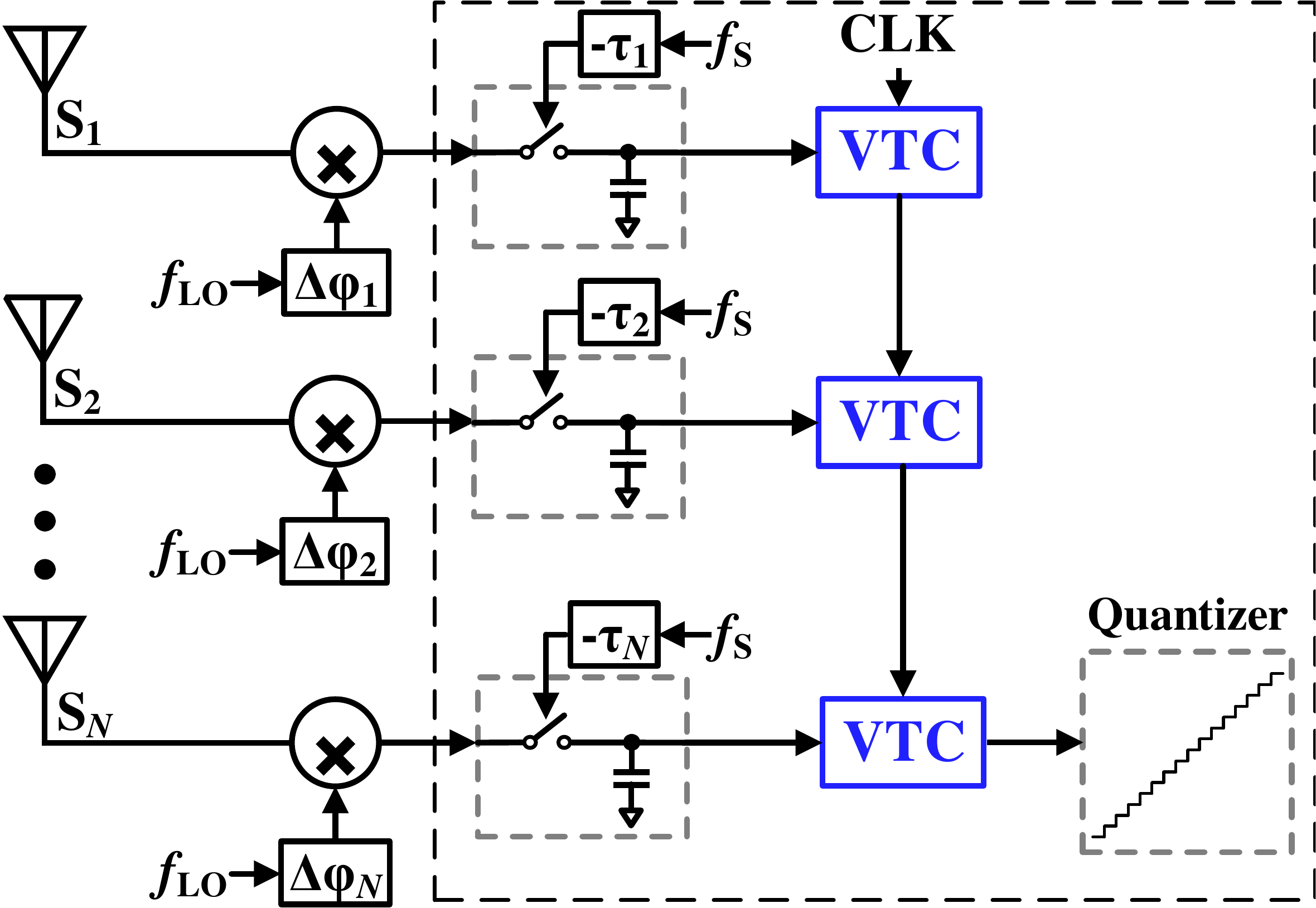}%
}\\
\vspace{-2mm}
\end{tabular}
\caption{(a)\label{fig:OTA_BF} Amplifier based summing architecture, and (b)  \label{fig:VTC_BF} Time domain pipeline summing architecture.} 
\vspace{-4mm}

\end{figure}

\subsubsection{Signal combiners} The signal combiner is a critical component for wideband discrete-time beamforming that requires careful design considerations of several parameters including gain, bandwidth, number of summing channels (each channel refers to a downconverted \ac{RF} signal from each antenna element), dynamic range, and power consumption. The multi-parameter optimization considering the required design specifications for each system can be addressed using either a closed-loop or open-loop summing amplifiers, shown in Fig.~\ref{fig:OTA_BF}(a). In \cite{ghaderi2019a}, a closed-loop \ac{OTA} is used whose design complexity and power consumption are scaled exponentially with the bandwidth. To support larger bandwidths and higher number of antennas, a closed loop integrator using ring amplifier \cite{hershberg-a, Hershberg2012} was demonstrated in \cite{Lin2021}. Ring amplifiers are technology scalable lending them good candidates to beamform signals from even larger number of antenna elements. In contrast to closed-loop amplifiers, summing can also be performed through daisy-chain linked \ac{VTC}s as shown in Fig.~\ref{fig:VTC_BF}(b). The serialized \ac{VTC}s each acting in open-loop is digitized by a multi-bit \ac{TDC} creating an equivalent of matrix-multiplying \ac{ADC}. 

\subsubsection{Precision clock generation}  
The required sampling phases of the discrete-time \ac{SSP} can be generated on-chip as illustrated in \cite{ghaderi2019a, Lin2021}. To achieve higher precision and matching following Nyquist criterion as well as use the same architecture for different \ac{SSP} modes, the clock generation requires \textit{N} low-power precision \ac{PI} and  time-interleavers \cite{ghaderi2019b,ghaderi2019a,Boljanovic2021}.  The time-interleaver output is applied to interleaved multiply-and-accumulate units that enables them to span the required delay range while meeting the Nyquist bandwidth. 

High precision linear \ac{PS}s are desired to meet the delay range and resolution requirements for the switched-capacitor array as discussed in Section~\ref{sec:syst:ttdarch}. Recent works have demonstrated \ac{PI}s  as digital-to-phase generator in different applications including clock-and-data recovery  \cite{hsieh2008,he2006}  and outphasing transmitters \cite{liu2009,su_gfsk_2010}. In general, several architectures have been proposed to implement \ac{PI} which generates output of phases of weighted sum of two input clock phases based on the digital input code.  

The underlying concept of \ac{PI} can be studied from \cite{weinlader_precision_2001} which describes the process as addition of two phase-shifted edges to produce a new edge with the transition in between them. 
The output is a superposition of two exponential curves formed by the merged driver output resistance and the capacitive load. 
The linearity of the \ac{PI} not only depends on the time difference between the input signals but also their rise times. It has been observed that the non-linearity for a step input response is more whereas it is less for the inputs with finite rise time inputs. 

\begin{figure}[t]
    \begin{center}
        \includegraphics[width=0.5\columnwidth]{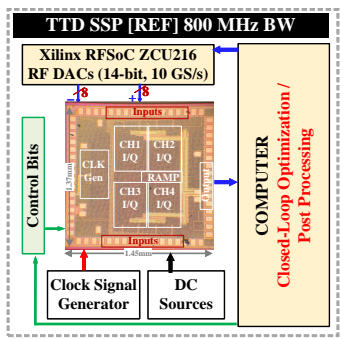}
    \end{center}
    \vspace{-2mm}
    \caption{Recent discrete-time \ac{TTD} \ac{SSP}s supporting 800MHz BW.  \cite{Lin2021}}
    \vspace{-4mm}
    \label{fig:TTD_testbed}
\end{figure}

The current-steering \ac{DAC} have been popular architectures for implementing \ac{PI} \cite{Bulzacchelli2006}. However, the linearity and resolution are greatly impacted by the \ac{DAC}. As mentioned above, to generate finite rise time at the input, a slew rate control buffer is implemented which requires additional power.  The summation of the direct in-phase and quadrature-phase however causes nonlinearity with respect to the codes. Alternatively, current-mode logic based \ac{PI} architectures can be adopted if the inputs are sinusoidal signals. Instead of interpolating between two consecutive quadrature phases, it can be done in multiple cascaded stages. Interpolation between signals with small phase differences ($<45^{\circ}$) is more linear than direct interpolation between the quadrature inputs. Though the current-mode logic based \ac{PI} has lower swing which consumes less dynamic power, it suffers from linearity issue where as the inverter-based \ac{PI}s suffers from high dynamic power. Additionally, the \ac{PI}’s output gets affected by the power supply variations. Low-dropout regulators can be applied between the main power supply and the \ac{PI} supply to alleviate this issue. As such, its important to consider the design trade-offs for $N$ \ac{PI}s specifically linearity and power consumption.


\subsection{Multi-antenna testbed and validation}
Another important step in the design process is developing the test interface between the \ac{mmW} front-end and the BB \ac{TTD} \ac{SSP} as well as the digital back-end. The wide modulated bandwidths and the large number of downconverted channels from \ac{mmW} to baseband due to both differential and quadrature configurations challenges the signal generation and post-processing. In addition, all the channels should be sychronized in time with picosecond accuracies to enable efficient data communications. Fig.~\ref{fig:TTD_testbed}  shows a sample closed-loop testbed  for 800~MHz \ac{TTD} \ac{SSP} recently demonstrated in \cite{Lin2021}. High speed \ac{DAC}s and \ac{ADC}s in ZCU RFSoC family make it possible to generate wideband signals from \ac{DAC}s and perform data capture from the high speed \ac{ADC} through the same graphic user interface control which serves both as an arbitrary waveform generator and spectrum analyzer in \cite{ghaderi2019a}. In \cite{Lin2021}, 800 MHz modulated bandwidth with 16 synchronized channels was generated and the outputs post-processed after digitization in MATLAB offline. 

\begin{figure}
    \begin{center}
        \includegraphics[width=0.485\textwidth]{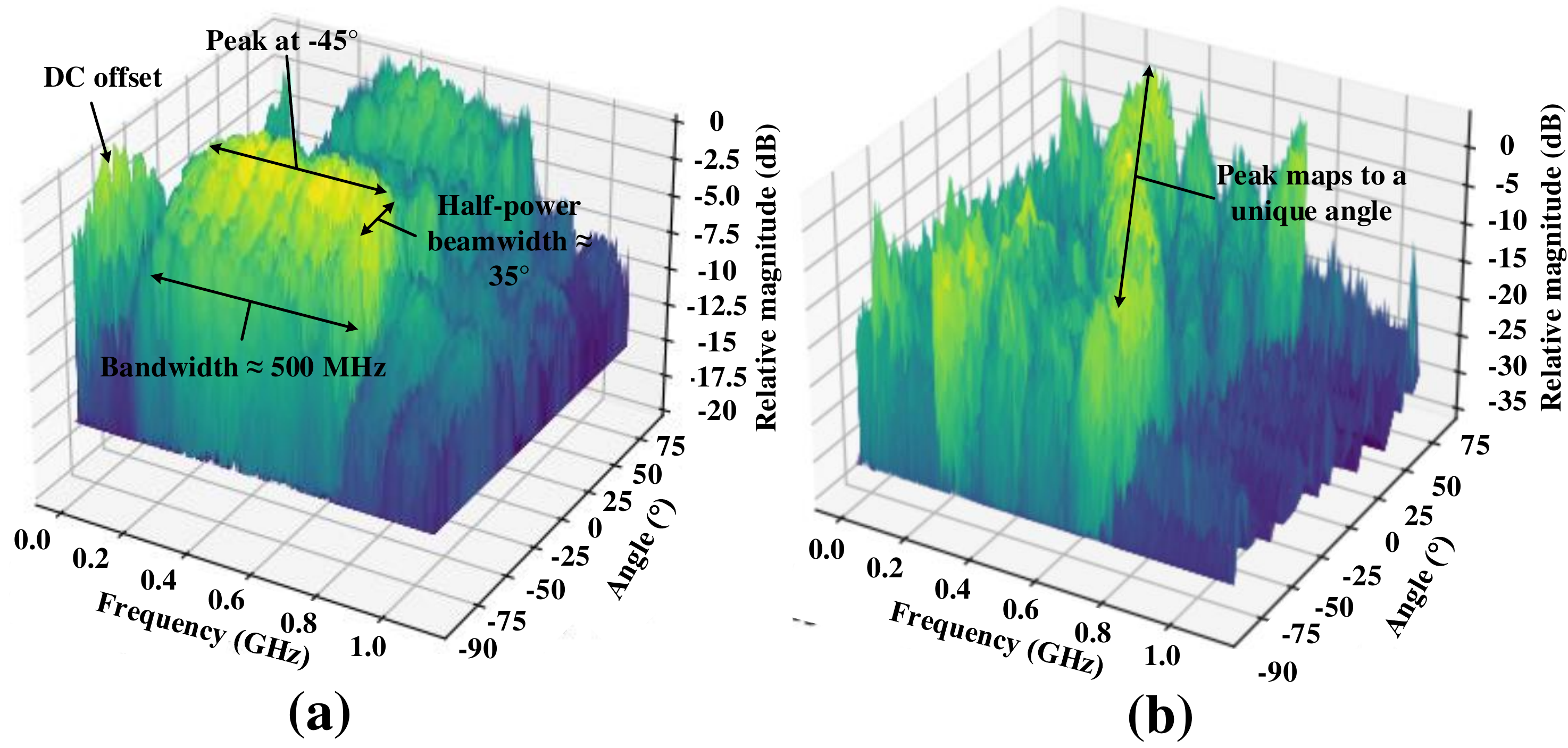}
    \end{center}
    \vspace{-2mm}
    \caption{Measured 3D plots of 4-element 500 MHz BW array for (a) data communications, and (b) rainbow beam-training.}
    \vspace{-4mm}
    \label{fig:ttd3dplot}
\end{figure}

Besides synchronization and large number of channels, repeated parameterized sweeps across frequencies and angles to characterize a multi-antenna transceiver undertakes significant efforts without closed-loop automation in place. In \cite{ghaderi2020}, the \ac{TTD} \ac{SSP} (in particular, \ac{TDC} and the time amplifiers) were optimized using particle swarm optimization techniques from \cite{Shrestha2020}. While the optimization was limited to \ac{TDC} only and not the entire \ac{SSP}, this method shows the potential opportunities to further extend the test bench automation for closed-loop signal path optimization. 

In their recent work, the authors have used computer vision techniques for testbed automation in \cite{lin_jssc_2021}. These techniques reduces the strenuous manual calibration and test process significantly and enables generation of data-intensive 3D plots (Fig.~\ref{fig:ttd3dplot}(a)) and frequency-angular maps (Fig.~\ref{fig:ttd3dplot}(b)) that can be used to create accurate dictionaries for calibration in data communications mode. The shown plots represent 72 individual configurations when the 4-element array is configured to -45$^\circ$ with 500 MHz modulated bandwidth at the input. The measured half-power beamwidth  is close to the theoretical estimated 35$^\circ$. Similar map can be generated for the rainbow beam-training as shown in Fig.~\ref{fig:ttd3dplot}(b) which shows the efficacy of the proposed algorithm. Additional measurement details for these plots are captured in \cite{lin_jssc_2021}. The automated test process further minimizes the human errors in the manual process. In addition, this enables benchmarking entire system from the antenna to the higher network layers combined with machine learning algorithms in future.


%
%

\section{Future Work}
\label{sec:futworks}

\subsection{3D rainbow beam with planar \ac{TTD} array}
\begin{figure}[t]
    \centering
	\vspace{0mm}
	\includegraphics[width=0.8\columnwidth]{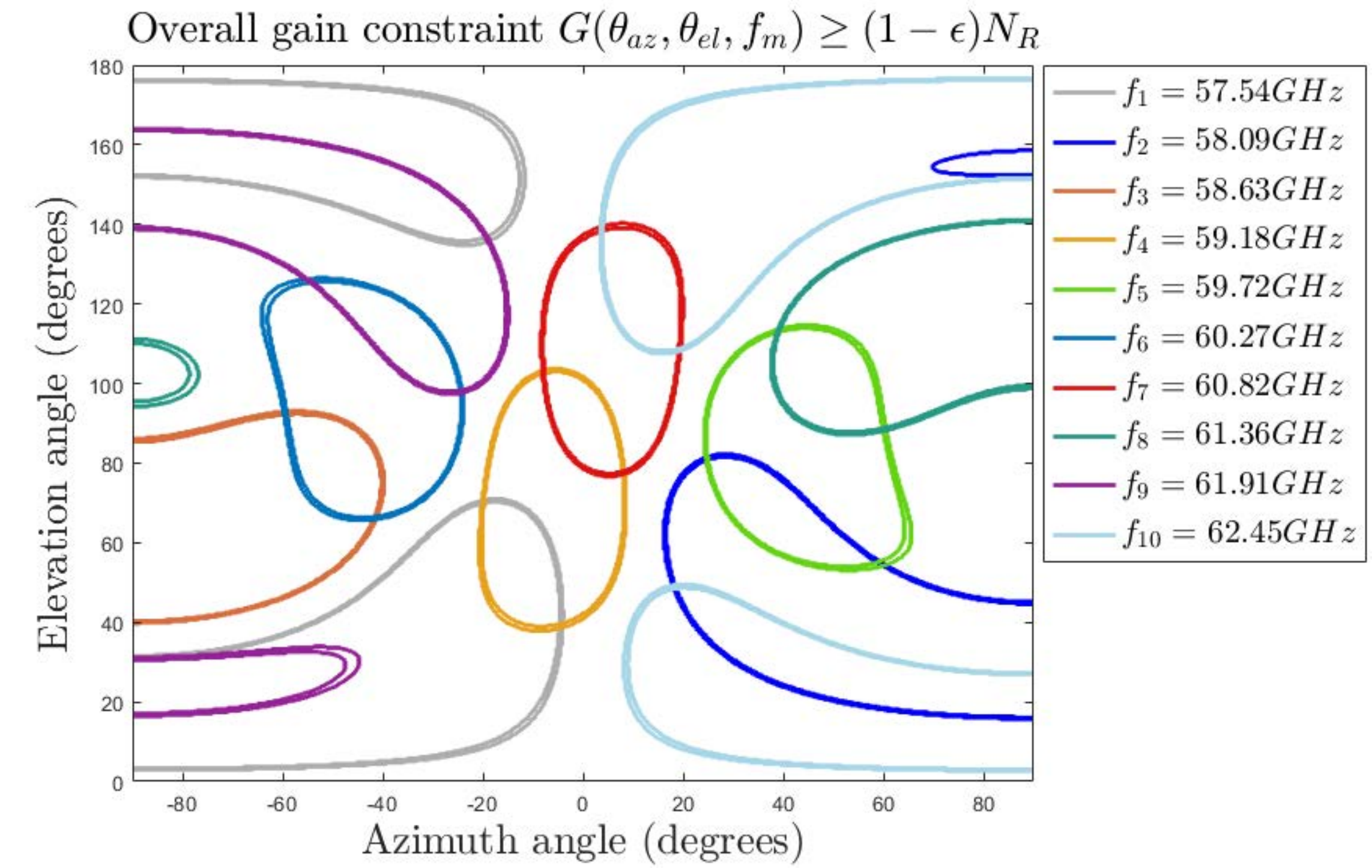}
	\vspace{-2mm}
	\caption{The simulated 3dB beam contour of frequency dependent steering from a planar \ac{TTD} array. $N_x=4$, $N_y=2$, $M=10$ subcarriers, $\Delta\tau_x = 1/\mathrm{BW}$; $\Delta\tau_y = 7/\mathrm{BW}$.}
	\vspace{-4mm}
	\label{fig:3d_rainbow_beam}
\end{figure}

In principle, the frequency dependent beam steering can be extended into the planar \ac{TTD} array. In our recent work \cite{TTD3d} we studied beamforming codebooks with the uniform delay steps $\Delta\tau_x, \Delta\tau_y$, introduced in the antenna elements in the x-y plane before combining, i.e., the received signal in the $(i,j)$-th element is delayed by $(i-1)\Delta\tau_x + (j-1)\Delta\tau_y$. We showed that uniform delay spacing can create 3D beams such that different frequency components are steered toward a unique spherical direction $(\theta,\phi)$. Further, the entire 3D angular space can be covered by at least one beam. Fig.~\ref{fig:3d_rainbow_beam} presents the frequency dispersing steering for 10 frequency components. However, this preliminary study also reveals a number of design challenges from both circuits and system perspectives. On the one hand, the required maximum delay range is increased. For planar array with a total $N$ elements, the required delay range is at least $2(N-1)/\mathrm{BW}$, a factor of 2 larger than its linear array counterpart. On the other hand, as compared to the rainbow beam using a linear \ac{TTD} array, the 3D rainbow beam using planar array has some unique characteristics. More specifically, beams from some frequency component experience a reduced gain as compared to others. There is non-trivial challenge to calibrate the beam pattern and design beam training algorithm tailored for this non-ideality. 

\begin{figure}[t]
    \centering
	\vspace{0mm}
	\includegraphics[width=0.35\textwidth]{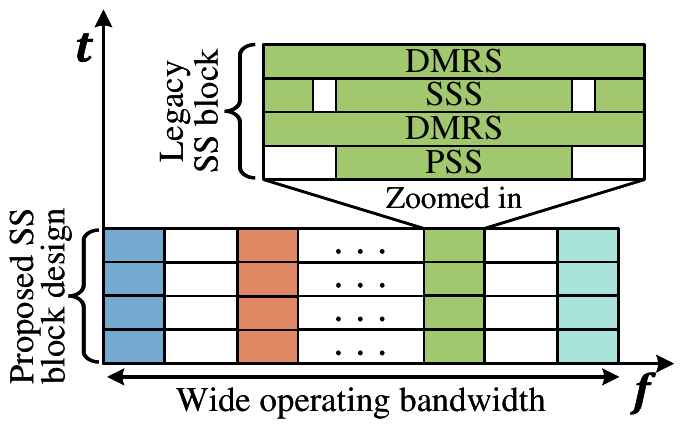}
	\vspace{-2mm}
	\caption{Illustration of the proposed \ac{SS} block design for \ac{TTD}-based beam training. At the \ac{UE} side, bandwidth parts of different colors are combined from different angular directions using a \ac{TTD} array.}
	\vspace{-4mm}
	\label{fig:ss_block}
\end{figure}

\subsection{\ac{TTD} rainbow beam enabled multiple access}
The beam training is not the only application of the \ac{TTD} array dispersive rainbow beam. In \cite{li2021rainbowlink}, a novel orthogonal frequency division multiple access scheme is proposed. Utilizing a \ac{TTD} array at the network infrastructure, different frequency resources are steered toward different unique directions to serve users. There are a number of advantages that make such scheme appealing for the future IoT application. Firstly, the hardware bandwidth of each user can be much smaller than the infrastructure, because each of them only access a portion of the frequency. This facilitates a power and cost friendly design for the terminals. Secondly, the conventional time division beam switching can be eliminated, because different users are simultaneously served using different frequency resource implicitly allocated by their corresponding rainbow beam direction. It further brings the potential to reduce the beam management overhead to support IoT applications with strict latency requirements \cite{li2021rainbowlink}. However, the challenge remains for the device to precisely synchronize with the network in frequency, because rainbow beam heavily relies on the mapping between frequency and beams. To this end, a frame structure and synchronization procedure that is compatible with existing standard is critical.

\subsection{Compatibility of \ac{TTD} Beam Training with \ac{5G} Standard}

Exhaustive beam sweeping with synchronization is the main part of the \ac{IA} procedure in \ac{5G} New Radio Standard. It is performed using periodic bursts of up to 64 \ac{SS} blocks, each associated with a different steering direction \cite{Zorzi:tutorial}. An \ac{SS} block is a group of 4 consecutive \ac{OFDM} symbols with 240 subcarriers ($20$ resource blocks) that carry the \ac{PSS}, \ac{SSS}, and \ac{DMRS}. The \ac{PSS} and \ac{SSS} are leveraged for synchronization between the \ac{BS} and \ac{UE}, while the \ac{DMRS} can be used to estimate the received power associated with that direction, i.e., \ac{SS} block.

Since the proposed \ac{TTD} beam training is fundamentally different from beam sweeping, its compatibility with the current \ac{5G} New Radio Standard becomes an important problem. In our preliminary work, we proposed a different design of the \ac{SS} block, which exploits wide available bandwidth at \ac{mmW} frequencies and supports \ac{TTD} beam training, as illustrated in Fig.~\ref{fig:ss_block}. In the proposed design, the existing reference signals are repeated multiple times across the operating bandwidth. With such training waveform, the \ac{UE} is capable of receiving at least one entire legacy \ac{SS} block using its frequency-dependent beam steering. The \ac{UE} can then leverage \ac{DSP} to extract the reference signals to achieve synchronization with the \ac{BS} and determine the best steering direction. The proposed \ac{SS} block design removes the need for \ac{SS} bursts and thus minimizes the required overhead in \ac{TTD} beam training. However, the design depends on different system parameters, including the number of subcarriers, number of \ac{UE} antennas, and resolution of \ac{UE} beam probing. We plan to address these practical design questions in our future work.

%
%

\section{Conclusions}
\label{sec:conclusions}
Reconfigurable \ac{TTD} arrays are essential for emerging \ac{mmW} wireless communications demanding wide bandwidths with ultra-low-latency in direction finding compared to current wireless standards. This magazine article presents a comprehensive overview of the recent developments in \ac{TTD} arrays enabling multiple \ac{SSP} functions. Delay compensation techniques are exploited to prevent beam squint during data communications while introducing intentional beam squint for fast beam-training. The article combines the beam-training algorithm with the underlying architectural and circuit design considerations. The exciting developments in \ac{TTD} arrays presented herein lays out a future path for enabling next-generation network and physical infrastructure wireless solutions on a large-scale with 3D direction finding for communications-on-the-move applications, massive machine type communications, and standardization in emerging wireless standards. 


%

\ifCLASSOPTIONcaptionsoff
  \newpage
\fi



%
%
%

\bibliographystyle{IEEEtran}
\end{document}